\documentclass[12pt]{article}
\usepackage{amssymb}
\usepackage{amsfonts}
\usepackage{mathrsfs}
\usepackage{amsmath}
\usepackage{slashed}
\usepackage{url}
\usepackage{bm}
\usepackage{upgreek}
\usepackage{hyperref}
\usepackage{graphicx}
\usepackage{setspace}
\usepackage[compact]{titlesec}
\usepackage[letterpaper,left=1.25in,right=1.25in,top=1in,bottom=1in,nohead]{geometry}
\usepackage{cite}
\usepackage{abstract}
\usepackage{lmodern}
\usepackage[T1]{fontenc}
\usepackage{xcolor}
\definecolor{dark-blue}{rgb}{0.15,0.15,0.4}

\hypersetup{
    colorlinks=true,       
    linkcolor=black,          
    citecolor=black,        
    urlcolor=dark-blue
}

\titleformat{\section}{\normalsize\bfseries}{\thesection}{1em}{}
\titleformat{\subsection}{\normalsize\bfseries}{\thesubsection}{1em}{}

\numberwithin{equation}{section}

\def\^{{\wedge}}
\def\*{{\star}}
\def\bar{\overline}

\def\e#1{{\rm e}^{\, #1}}
\def\wt#1{\widetilde#1}
\def\ha{\frac{1}{2}}

\def\Hol{{\mathop{\rm Hol}}}

\def\Vol{\mathop{\rm Vol}}
\def\Map{{\mathop{\rm Map}}}

\def\Ker{{\mathop{\rm Ker}}}
\def\Im{\mathop{\rm Im}}
\def\Pic{{\mathop{\rm Pic}}}

\def\mod{\mathop{\rm mod}}

\def\vol{{\mathop{\rm vol}}}
\def\Vol{{\mathop{\rm Vol}}}

\def\Tr{{{\mathop{\rm Tr}}}}

\def\BC{{\mathbb C}}

\def\BL{{\mathbb L}}
\def\BP{{\mathbb P}}

\def\BR{{\mathbb R}}

\def\BZ{{\mathbb Z}}

\def\CA{{\mathcal A}}

\def\CD{{\mathcal D}}

\def\CF{{\mathcal F}}
\def\CG{{\mathcal G}}
\def\CH{{\mathcal H}}

\def\CL{{\mathcal L}}

\def\CN{{\mathcal N}}
\def\CO{{\mathcal O}}

\def\CX{{\mathcal X}}

\def\Fe{{\mathfrak e}}

\def\SH{{\mathscr H}}
\def\SJ{{\mathscr J}}
\def\SL{{\mathsf L}}
\def\SO{{\mathsf O}}
\def\SQ{{\mathsf Q}}

\def\SW{{\mathsf W}}
\def\SV{{\mathsf V}}

\newcommand{\version}{\normalfont May 2014}

\begin{document}
\begin{titlepage}
\begin{flushright}
{\small\ttfamily arXiv:1405.2483}
\end{flushright}
\begin{center}
\vspace{2cm}
{\large\bfseries  Abelian Duality at Higher Genus}\\
\vspace{1cm}
Chris Beasley\\
\vspace{3mm}
{\small\sl Department of Mathematics, Northeastern University, Boston, MA 02115}
\vspace{-5mm}
\end{center}
\begin{abstract}
\baselineskip=18pt
In three dimensions, a free, periodic scalar field is related by
duality to an abelian gauge field.  Here I explore aspects of
this duality when both theories are quantized on a Riemann surface of
genus $g$.  At higher genus, duality involves an identification of winding
with momentum on the Jacobian variety of the Riemann surface.  I also
consider duality for monopole and loop operators on the surface and
exhibit the operator algebra, a refinement of the Wilson-\mbox{'t
  Hooft} algebra.
\end{abstract} 
\vfill\hskip 1cm\version
\end{titlepage}
\begin{onehalfspace}
\tableofcontents\noindent\hrulefill
\section{Introduction}\label{Intro}

Very broadly, dualities in quantum field theory often involve an
interchange between classical and quantum data.  Perhaps the
simplest and best-known example occurs for the theory of a free
periodic\footnote{``Periodic'' is perhaps better stated as ``circle-valued''.} scalar field ${\phi \sim \phi + 2\pi}$ on a Riemann surface
$\Sigma$, with sigma model action 
\begin{equation}\label{SIGMAI}
{\bf I}(\phi) \,=\, \frac{R^2}{4\pi} \int_\Sigma d\phi\^\*d\phi \,=\,
\frac{R^2}{4\pi} \int_\Sigma d^2x \, \sqrt{h} \, \partial_\mu\phi
\, \partial^\mu\phi\,,\qquad\qquad \mu=1,2\,.
\end{equation}
Here $R$ is a parameter which determines the radius of the
circle for maps ${\phi:\Sigma\to S^1}$, and $\*$ is the Hodge
star associated to a given metric $h$ on $\Sigma$.

When $\phi$ is quantized on the circle, meaning that we take ${\Sigma = \BR 
  \times S^1}$, one finds that the Hilbert space is graded
by a pair of integers ${\left(p, w\right)}$,
\begin{equation}\label{DSUM}
\SH_{S^1}^{} = \bigoplus_{\left(p,w\right) \in \BZ\oplus\BZ} \SH_{S^1}^{\,p,w}\,.
\end{equation}
The integers $p$ and $w$ are naturally interpreted as charges for a
combined $U(1)_\ell\times U(1)_r$ action on $\SH_{S^1}$, under which each
summand in \eqref{DSUM} transforms with the specified weights.
The integer $p$ is associated to the global $U(1)_\ell$ symmetry under
which the value of $\phi$ shifts by a constant,
\begin{equation}\label{UONEL}
U(1)_\ell:\quad \phi\,\longmapsto\, \phi \,+\, c\,,\qquad\qquad c\,\in\,\BR/2\pi\BZ\,.
\end{equation}
This transformation clearly preserves the classical action in
\eqref{SIGMAI}.  Concretely, $p$ labels the states in $\SH_{S^1}$
which arise from the quantization of the constant mode ${\phi_0 \in
  S^1}$ of the scalar field.  These states correspond to a
Fourier basis for $L^2(S^1;\BC)$ as below,
\begin{equation}
\Psi_p(\phi_0) \,=\, \e{i\,p\,\phi_0}\,,\qquad\qquad p\,\in\,\BZ\,,
\end{equation}
and $p$ is the momentum conjugate to $\phi_0$.

The other charge $w$ describes the winding-number of $\phi$ as a map
from the circle to itself.  Hence $w$ labels the connected components of
the configuration space 
\begin{equation}\label{BIGCONFX}
\CX \,=\, \bigsqcup_{w\in\BZ} \, \CX_w\,,\qquad\qquad\CX \,=\,
\Map(S^1,\,S^1)\,,
\end{equation}
where
\begin{equation}
\CX_w \,=\, \big\{\phi:S^1\to S^1\,\big|\,\phi(x+2\pi) \,=\,
  \phi(x) \,+\, 2\pi w\big\}\,.
\end{equation}
Each component of the classical configuration space must be
quantized separately, and those states which arise from $\CX_w$ span the
subspace of the Hilbert space at the grade $w$.

Though the Hilbert space on $S^1$ is bigraded by
$(p,w)\in\BZ\oplus\BZ$, the individual gradings have very
different physical origins.  The momentum 
$p$ appears only after quantization, so the grading by $p$ is inherently
quantum.  Conversely, the grading by winding-number $w$ can be
understood in terms of the topology of the configuration space
$\CX$, so the grading by $w$ is classical. 

In a similar vein, the conserved currents on $\Sigma$ associated to
the  ${U(1)_\ell \times U(1)_r}$ global symmetry are respectively
\begin{equation}\label{CURRJS}
j_\ell \,=\, d\phi\,,\qquad\qquad j_r \,=\, \*d\phi\,.
\end{equation}
The topological current $j_r$, whose charge is the winding-number $w$,
trivially satisfies the conservation 
equation ${d{}^{\dagger} j_r  = 0}$ (with ${d{}^\dagger = -\* d \*}$)
for arbitrary configurations of the field $\phi$ on $\Sigma$.  By
contrast, ${d{}^\dagger j_\ell = 
  0}$ only when $\phi$ satisfies the classical equation of motion
${\triangle \phi = d{}^\dagger d\phi = 0}$.  Thus conservation of
$j_\ell$ is a feature of the dynamics -- or lack thereof -- in the
abelian sigma model.

Because $\Sigma$ has dimension two, the conserved currents $j_\ell$
and $j_r$ are exchanged under the action by Poincar\'e-Hodge duality
on the space of one-forms $\Omega^1_\Sigma$.  As familiar, the
classical action by Poincar\'e duality extends to an action by \mbox{T-duality}
\cite{Buscher:1985,BuscherSK,BuscherQJ,Giveon:1994fu,RocekPS} on the
quantum field $\phi$, under which the respective quantum and classical
gradings by momentum and winding are exchanged, and the parameter
$R$ in \eqref{SIGMAI} is inverted to $1/R$.  See for instance Lecture
8 in \cite{Deligne} for further discussion of abelian duality on
$\Sigma$.

The present paper is a continuation of \cite{BeasleyI:2014}, in which
we examine global issues surrounding abelian duality in dimension
three, on a Riemannian three-manifold $M$.  In this case, duality now 
relates \cite{BrodaWG,Kapustin:2009av,ProdanovJY,Deligne} the periodic
scalar field ${\phi:M\to S^1}$ to a $U(1)$ gauge 
field $A$ on $M$.  Both quantum field theories 
are free, so both can be quantized on the product ${M =
  \BR \times \Sigma}$ to produce respective Hilbert spaces $\SH_\Sigma^{}$ and
$\SH_\Sigma^\vee$.  Duality is an equivalence of quantum field 
theories, so we expect an isomorphism\footnote{Our notation for the
  Maxwell Hilbert space $\SH_\Sigma^\vee$ is {\em not} intended to suggest that
 it is naturally dual as a vector space to the scalar Hilbert space $\SH_\Sigma$.}
\begin{equation}\label{ISOM}
\SH_\Sigma^{} \,\simeq\, \SH^\vee_\Sigma\,.
\end{equation}
Exploring how the identification in \eqref{ISOM} works when 
$\Sigma$ is a compact Riemann surface of genus $g$ will be our main
goal in this paper.

Just as T-duality acts in a non-trivial way on $\SH_{S^1}$ by
exchanging momentum and winding in \eqref{DSUM}, we will see that the dual
identification ${\SH_\Sigma^{} \simeq \SH^\vee_\Sigma}$ relies upon an
analogous exchange of classical and quantum data for the
scalar field and the gauge field on $\Sigma$.  However, the Hilbert spaces
$\SH_\Sigma^{}$ and $\SH^\vee_\Sigma$ are now more interesting due to 
their dependence on the geometry of the Riemann surface $\Sigma$, and
unraveling the isomorphism in \eqref{ISOM} turns out to be a richer
story than for quantization on $S^1$. 

In genus zero, when ${\Sigma = \BC\BP^1}$ and the Hilbert space has a
physical interpretation via radial quantization on $\BR^3$, nothing
that we say will be new.  As usual in the world of Riemann
surfaces, though, genus zero is a rather degenerate case, and several 
important features only emerge at genus ${g \ge 1}$.  From the
perspective of the scalar field, these features are related to
topological winding-modes on $\Sigma$, and from the perspective of the gauge
field, they are related to the existence of a moduli space of
non-trivial flat connections on $\Sigma$.

Given the current rudimentary understanding of duality, especially in
dimension three, the existence of any tractable example is
important.  Both this work and \cite{BeasleyI:2014} are motivated by
questions about non-abelian duality for a certain topological version
of the ${\CN=8}$ supersymmetric Yang-Mills theory in three dimensions,
considered to a certain extent in \S $3.3$ of \cite{Witten:2009mh}.
From the latter perspective, the abelian analysis here provides a
useful toy model in which everything can be understood directly and
in detail.

\medskip\noindent{\sl The Plan of the Paper}\medskip

In Section \ref{Hilbert}, we construct the respective Hilbert spaces
$\SH_\Sigma^{}$ and $\SH^\vee_\Sigma$ associated to the periodic
scalar field and the abelian gauge field on ${M = \BR \times
  \Sigma}$.  Because the quantum field theories are free, the 
quantization holds no mystery and can be carried out quite rigorously,
if one wishes.  Both Hilbert spaces depend on the detailed
choice of the Riemannian metric on $\Sigma$.  In either case, though, 
we identify a particularly simple, infinite-dimensional subspace of
`quasi-topological' states which depend only upon the overall volume
and complex structure of $\Sigma$.  These quasi-topological states are
exchanged under duality, analogous to the exchange of momentum and
winding states for quantization on $S^1$.

In Section \ref{Algebra}, we proceed to the consider the algebra
satisfied by a natural set of operators (Wilson loops, vortex
loops, and monopole operators in the language of Maxwell theory) which
act on the Hilbert spaces constructed in Section $2$.  For a free
quantum field theory, there is only one possible operator algebra that
can arise --- namely, the Heisenberg algebra, in a suitable geometric
realization.  When ${\Sigma = \BC\BP^1}$ there is not much to 
say, but in higher genus, the operator algebra has a non-trivial
holomorphic dependence on $\Sigma$ that seems not to have been
previously noted.  This algebra is a refinement of the celebrated
Wilson-\mbox{'t Hooft} algebra \cite{tHooft:1977hy}.  See 
\cite{Freed:2006ya,Freed:2006yc} for a somewhat related appearance of
the Heisenberg algebra in four-dimensional Maxwell theory, and
\cite{Borokhov:2002ib,Borokhov:2002cg,Drukker:2012sr,Intriligator:2013lca,Kapustin:2012iw}
for some recent discussions of monopoles and vortices in the setting
of ${\CN=2}$ supersymmetric gauge theory.

Along the way, we also consider in Section \ref{Monopoles} the dual
identification of operators acting on ${\SH_\Sigma^{} \,\simeq\,
  \SH^\vee_\Sigma}$.  For the convenience of the reader, a
complementary review of the path integral perspective on the
order-disorder correspondence for our operators can be found in Section
$5.2$ of \cite{BeasleyI:2014}.

\medskip
\section{Abelian Duality on a Riemann Surface}\label{Hilbert}

In this section, we quantize both the periodic scalar field and the
abelian gauge field on a compact surface $\Sigma$ of genus $g$, with
Riemannian metric $h$.  We then compare the results.

\medskip\noindent{\sl Some Geometric Preliminaries}\medskip

Though the quantum field theories under consideration are free, they
definitely depend upon the choice of the metric $h$ on $\Sigma$.  The
most basic invariant of $h$ is the total volume 
\begin{equation}\label{VOLSIG}
\ell^2 \,=\, \int_\Sigma \, \vol_\Sigma\,,\qquad\qquad \vol_\Sigma
\,=\, \*1 \,\in\,\Omega^2_\Sigma\,,
\end{equation}
where $\ell$ is the length associated to the chosen metric.  

As in \cite{BeasleyI:2014}, the Hamiltonians for both the 
scalar field and the gauge field on $\Sigma$ will depend upon a
parameter $e^2$, identified with the electric coupling in the Maxwell
theory on ${M = \BR \times \Sigma}$.  All our constructions will
respect the classical scaling under which the metric $h$ transforms by 
\begin{equation}\label{SCALEh}
h \,\longmapsto\, \Lambda^2\,h\,,\qquad\qquad \Lambda\,\in\,\BR_+\,,
\end{equation}
along with 
\begin{equation}
\ell \,\longmapsto\, \Lambda\,\ell\,,
\end{equation}
and 
\begin{equation}\label{SCALEe}
e^2 \,\longmapsto\, \Lambda^{-1} \, e^2\,.
\end{equation}
Hence $\ell$ and $e^2$ are redundant parameters, since either can be
scaled to unity with an appropriate choice of $\Lambda$ in \eqref{SCALEh}.
Nonetheless, we leave the dependence on both $\ell$ and $e^2$
explicit, so that the naive dimensional analysis holds.

At least when the genus of $\Sigma$ is positive, a more refined
invariant of the metric $h$ is the induced complex 
structure on $\Sigma$.  Concretely, specifying a complex structure on
$\Sigma$ amounts to specifying a Hodge decomposition for complex one-forms
\begin{equation}\label{HODGE}
\Omega^1_\Sigma\otimes\BC \,\simeq\, \Omega^{1,0}_\Sigma \!\oplus
\Omega^{0,1}_\Sigma\,,
\end{equation}
where $\Omega^{1,0}_\Sigma$ and $\Omega^{0,1}_\Sigma$ refer to complex
one-forms of given holomorphic/anti-holomorphic type.  With this
decomposition, one can define a Dolbeault operator $\bar\partial$
by projection onto $\Omega^{0,1}_\Sigma$, from which one obtains a
notion of holomorphy on $\Sigma$.

The Hodge star associated to the metric $h$ satisfies ${\*^2 = -1}$
when acting on $\Omega^1_\Sigma$.  The
eigenspaces of the Hodge star then provide the decomposition in
\eqref{HODGE}, where by convention 
\begin{equation}\label{HODGEII}
\* \,=\, -i \,\,\hbox{ on }\,\,\Omega^{1,0}_\Sigma\,,\qquad\qquad\qquad
\*\,=\,+i\,\,\hbox{ on }\,\,\Omega^{0,1}_\Sigma\,.
\end{equation}
In this manner, the metric $h$ determines a complex structure on $\Sigma$.

Finally, we will make great use of the de Rham Laplacians $\triangle_0$ and
$\triangle_1$ acting on differential forms of degrees zero and one on
$\Sigma$.  As usual, both Laplacians are defined in terms of the $L^2$-adjoint $d{}^\dagger$ via ${\triangle_{0,1} = d{}^\dagger d
  + d d{}^\dagger}$.  With this convention, the Laplacian is a positive
operator.  Because $\Sigma$ is smooth and compact, the spectra of
$\triangle_0$ and $\triangle_1$ are discrete, and the kernels of
each Laplacian are identified with the cohomology groups 
\begin{equation}
H^0(\Sigma;\BR) \,=\, \BR\,,\qquad\qquad H^1(\Sigma;\BR) \,=\, \BR^{2 g}\,.
\end{equation}

\subsection{The Quantum Sigma Model}\label{Scalar}

As in Section $2$ of \cite{BeasleyI:2014}, we consider a periodic
scalar field $\phi$ on ${M = \BR \times \Sigma}$,
\begin{equation}
\phi:M \,\longrightarrow\, S^1 \,\simeq\,\BR/2\pi\BZ\,,
\end{equation}
where we interpret $\phi$ as an angular quantity, subject to the
identification 
\begin{equation}
\phi\,\sim\,\phi \,+\, 2\pi\,.
\end{equation}
Unlike in \cite{BeasleyI:2014}, though, for the purpose of
quantization we work in Lorentz signature $(-++)$ on ${\BR \times
  \Sigma}$, with the product metric
\begin{equation}\label{3DM}
ds^2_M \,=\, -dt^2 \,+\, h_{z \bar z} \, dz \!\otimes\! d\bar z\,.
\end{equation}
Here $t$ is interpreted as the ``time'' along $\BR$, and $(z,\bar z)$
are local holomorphic/anti-holomorphic coordinates on $\Sigma$.
Philosophically, quantization is more naturally carried out in
Lorentz as opposed to Euclidean signature, since only in the
former case does one expect to obtain a physically-sensible, unitary
quantum field theory.

In Lorentz signature on ${M = \BR \times \Sigma}$, the free sigma
model action is given by 
\begin{equation}\label{SIGMA0}
{\bf I}_0(\phi) \,=\, \frac{e^2}{4\pi} \int_{\BR \times \Sigma} \!\!dt\,
\Big[\left(\partial_t\phi\right)^2 \, \vol_\Sigma \,-\,
  d\phi\^\*d\phi\Big].
\end{equation}
Throughout, we follow the convention that the de Rham operator $d$ and the
Hodge star ${\star \equiv \star_\Sigma}$ refer to quantities on $\Sigma$, as opposed to
$M$.  As in \eqref{VOLSIG}, $\vol_\Sigma$ is the Riemannian volume
form on $\Sigma$ induced by the metric $h$.  Finally, $e^2$ is a
dimensionful parameter which will eventually be identified under
duality with the electric coupling in Maxwell theory on $M$.  

Under the scaling by $\Lambda$ in \eqref{SCALEh} and \eqref{SCALEe},
the volume form on $\Sigma$ transforms by 
\begin{equation}
\vol_\Sigma \,\longrightarrow\, \Lambda^2 \, \vol_\Sigma\,,
\end{equation}
and ${d\phi \^ \*d\phi}$ is invariant, a fact familiar in the context of
two-dimensional conformal field theory. Hence ${\bf I}_0(\phi)$ will
be invariant under \eqref{SCALEh} and \eqref{SCALEe} provided that the
time $t$ is also scaled by 
\begin{equation}\label{SCALEt}
t \,\longmapsto\, \Lambda \, t\,.
\end{equation}
This scaling of the time coordinate is moreover necessary for a homogeneous
scaling of the three-dimensional metric \eqref{3DM} on $M$.

As we observed in \cite{BeasleyI:2014}, the free sigma
model action in \eqref{SIGMA0} can be extended by topological
terms 
\begin{equation}\label{TOPI}
{\bf I}_1(\phi) \,=\, \frac{e^2}{2\pi} \int_{\BR \times
  \Sigma}\!\!dt\,\upalpha\^d\phi \,+\, \frac{\theta}{2\pi \ell^2}
\int_{\BR \times \Sigma}\!\!dt\,\,\partial_t\phi\cdot\vol_\Sigma\,.
\end{equation}
Here $\upalpha$ is a harmonic one-form on $\Sigma$,
\begin{equation}\label{ALPHP}
\upalpha \,\in\,\CH^1(\Sigma)\,,
\end{equation}
and $\theta$ is a real constant,
\begin{equation}
\theta \,\in\,\BR\,.
\end{equation}
Together, $\upalpha$ and $\theta$ specify the components of
the complex harmonic two-form $\upgamma$ that appears on the compact
three-manifold $M$ in \cite{BeasleyI:2014}.  The prefactor of $1/2\pi$
in \eqref{TOPI} is just a convention, and the factors of $e^2$ and
$1/\ell^2$ in the respective terms are dictated by invariance under
the scaling in \eqref{SCALEh}, \eqref{SCALEe}, and \eqref{SCALEt}.
Note also that the two-form ${\vol_\Sigma/\ell^2}$ which enters the
second term in \eqref{TOPI} is properly normalized to serve
as an integral generator for $H^2(\Sigma;\BZ)$.

We take the total sigma model action to be the sum
\begin{equation}\label{TOTSI}
\begin{aligned}
{\bf I}_{\rm tot}(\phi) \,&=\, {\bf I}_0(\phi) \,+\, {\bf
  I}_1(\phi)\,,
\end{aligned}
\end{equation}
or more explicitly,
\begin{equation}\label{TOTSII}
{\bf I}_{\rm tot}(\phi) \,=\, \frac{e^2}{4\pi} \int_{\BR \times \Sigma} \!\!dt\,
\Big[\left(\partial_t\phi\right)^2 \vol_\Sigma \,+\, 2 \, \frac{\theta}{e^2\ell^2} \, \partial_t\phi \, \vol_\Sigma \,-\,
  d\phi\^\*d\phi \,+\, 2 \, \upalpha\^d\phi\Big].
\end{equation}
Because $\upalpha$ is closed by assumption, ${d\upalpha = 0}$, the
topological terms in \eqref{TOTSII} do not alter the 
classical equation of motion 
\begin{equation}\label{EOFMP}
\partial_t^2 \phi \,+\, \triangle_0\phi \,=\, 0\,,
\end{equation}
a version of the usual wave equation on ${\BR \times \Sigma}$.
As a small check on our signs, note that  ${\triangle_0 \ge
  0}$ is positive while ${\partial_t^2
  \le 0}$ is negative, so the equation of motion
for $\phi$ does admit non-trivial, time-dependent solutions.  Clearly,
$\upalpha$ in \eqref{TOTSII} serves to distinguish the various
topological winding-sectors associated to the circle-valued map $\phi$.

Like $\upalpha$, the constant $\theta$ multiplies a term in the action
\eqref{TOPI} which is a total derivative.  Hence $\theta$ has no
effect on the classical physics.  However, $\theta$ does change 
the definition of the canonical momentum $\Pi_\phi$ conjugate to $\phi$,
\begin{equation}\label{MOMENTUM}
\Pi_\phi \,=\, \frac{e^2}{2\pi}\left(\partial_t\phi \,+\,
  \frac{\theta}{e^2 \ell^2}\right),
\end{equation}
in terms of which we write the classical Hamiltonian 
\begin{equation}\label{CLASSH}
{\bf H} \,=\, \int_\Sigma
\left[\frac{\pi}{e^2}\!\left(\Pi_\phi\,-\,\frac{\theta}{2\pi \ell^2}\right)^2 
  \!\vol_\Sigma \,+\, \frac{e^2}{4\pi}\, d\phi\^\*d\phi \,-\,
  \frac{e^2}{2\pi}\,\upalpha\^d\phi\right]\,.
\end{equation}
As will be clear, the quantum sigma model does depend upon $\theta$
non-trivially, and ${\theta\in\BR/2\pi\BZ}$ becomes an angular
parameter closely analogous to the theta-angle of Yang-Mills theory in two
and four dimensions.

In addition to the Hamiltonian ${\bf H}$, another important quantity 
is the conserved momentum ${\bf P}$ associated to the global $U(1)$
symmetry under which the value of $\phi$ shifts by a constant, exactly
as in \eqref{UONEL}.  Constant shifts in $\phi$ manifestly preserve
the sigma model action, with conserved current ${j = e^2\,d\phi/2\pi}$
and charge 
\begin{equation}\label{BIGP}
{\bf P} \,=\, \frac{e^2}{2\pi}\int_\Sigma \partial_t\phi \cdot
\vol_\Sigma\,.
\end{equation}
Because of the $U(1)$ symmetry, the Hilbert space for
$\phi$ will automatically carry an integral grading by the eigenvalue
of ${\bf P}$ when we quantize.

\medskip\noindent{\sl Classical Mode Expansion}\medskip

In principle, the Hilbert space for the periodic scalar field on the
surface $\Sigma$ is straightforward to describe, though the 
detailed spectrum of the Hamiltonian depends very much on the geometry
of $\Sigma$.

Very briefly, just as for quantization on $S^1$, the quantization
on $\Sigma$ involves a countable number of topological sectors,
corresponding to homotopy classes of the map ${\phi:\Sigma \to S^1}$.
These homotopy classes are labelled by a winding-number $\omega$
which is valued in the cohomology lattice 
\begin{equation}
\BL \,=\, H^1(\Sigma;\BZ) \,\simeq\, \BZ^{2 g}\,,
\end{equation}
as discussed for instance in Section $2.1$ of \cite{BeasleyI:2014}.  Abusing 
notation slightly, I write the cohomology class associated to 
the circle-valued map $\phi$ as 
\begin{equation}
\omega \,=\, \left[\frac{d\phi}{2\pi}\right] \,\in\, H^1(\Sigma;\BZ)\,.
\end{equation}
Globally, the configuration space ${\CX = \Map(\Sigma,S^1)}$
is a union of components 
\begin{equation}\label{COMPONENT}
\CX \,=\, \bigsqcup_{\omega \in \BL}\,\CX_\omega\,,
\end{equation}
where 
\begin{equation}
\CX_\omega \,=\, \left\{ \phi:\Sigma\,\to\,
  S^1\,\Biggr|\,\left[\frac{d\phi}{2\pi}\right] = \omega\right\}\,.
\end{equation}
As standard in quantum field theory, each component
${\CX_\omega\subset\CX}$ must be quantized separately, leading to a
topological grading by the cohomology lattice $\BL$ on the full Hilbert space
$\SH_\Sigma$.  This grading by ${\omega\in\BL}$ is the obvious
counterpart for quantization on $\Sigma$ to the grading \eqref{DSUM}
by winding-number ${w\in\BZ}$ for quantization on $S^1$.

Concretely, as the first step towards constructing the sigma model
Hilbert space, we solve the classical equation of motion for $\phi$ in
\eqref{EOFMP}.  For the moment, we assume $\phi$ to have trivial
winding, so that the time-dependent field ${\phi:\BR\times \Sigma\to
  S^1}$ can be equivalently considered as a map ${\phi:\BR \to \CX_0}$
to the identity component of $\CX$.  

The general solution of \eqref{EOFMP} then takes the form 
\begin{equation}\label{EIGEN}
\phi(t,z,\bar z) \,=\, \frac{\phi_0}{e^2 \ell} \,+\, \frac{2\pi t}{\ell}\,p_0
\,+\, \sqrt{\pi}\,\sum_{\lambda > 0} 
\frac{e^2}{\lambda}\,\psi_\lambda(z,\bar z) 
\,\Big[a_\lambda \, \e{-i\,\lambda\,t} \,+\, a_\lambda{}^{\!\dagger} \,
\e{i\,\lambda\,t}\Big].
\end{equation} 
Generalizing the standard solution to the wave equation on 
$\BR^{1,2}$, this expression for $\phi$ is written in terms of an
orthonormal basis $\left\{\psi_\lambda\right\}$ of eigenmodes for the
scalar Laplacian $\triangle_0$ on $\Sigma$, 
\begin{equation}\label{PSILAM}
\triangle_0 \psi_\lambda \,=\, \lambda^2 \,
\psi_\lambda\,,\qquad\qquad \psi_\lambda\,\in\,\Omega^0_\Sigma\,,
\end{equation}
with the convention that ${\lambda > 0}$ is a positive real number.
For simplicity, we assume that all non-vanishing eigenvalues
of the scalar Laplacian are distinct, so that each eigenfunction
$\psi_\lambda(z,\bar z)$ is uniquely labelled by $\lambda$.  Of
course, the precise spectrum for the scalar Laplacian depends
sensitively on the geometry of the surface $\Sigma$, but we will not
require any detailed information about the spectrum here, other than
that it is discrete.

To ensure invariance of the eigenmode expansion for $\phi$ under the
scaling in \eqref{SCALEh} and \eqref{SCALEe}, we employ the
invariant (and coupling-dependent) normalization condition 
\begin{equation}\label{NORM}
||\psi_\lambda||^2 \,=\, e^4 \int_\Sigma \psi_\lambda^2 \,\, \vol_\Sigma
\,=\, 1\,.
\end{equation}
A similar coupling-dependent normalization condition is used in $(2.46)$ of 
\cite{BeasleyI:2014}, for the same reason.
Accordingly, the constant function with unit norm on
$\Sigma$ is 
\begin{equation}
\psi_0 \,=\, \frac{1}{e^2\ell}\,.
\end{equation}
This constant function appears implicitly in \eqref{EIGEN} as
the coefficient of the zero-mode $\phi_0$.  Because $\phi$ is an
angular quantity with period $2\pi$, the zero-mode $\phi_0$ must have
its own periodicity 
\begin{equation}\label{RADO}
\phi_0 \,\sim\, \phi_0 \,+\, 2\pi e^2\ell\,.
\end{equation}
Hence $\phi_0$ effectively decompactifies in the large-volume limit
${\ell\to\infty}$ with $e^2$ fixed.

Otherwise, $p_0$, $a_\lambda$, and $a_\lambda{}^{\!\dagger}$ for
${\lambda > 0}$ in
\eqref{EIGEN} are constants which specify the classical solution for
$\phi$.  The constant ${p_0\in\BR}$ is real and determines the
classical momentum via 
\begin{equation}\label{BIGPp}
{\bf P} \,=\, \frac{e^2}{2\pi} \int_\Sigma \partial_t\phi \cdot
\vol_\Sigma \,=\, e^2 \ell \, p_0\,,
\end{equation}
whereas $(a_\lambda,a_\lambda{}^{\!\dagger})$ are a
conjugate pair of complex numbers associated to the oscillating modes
of $\phi$.  The various factors of $e^2$ and $\ell$ sprinkled about
\eqref{EIGEN} are necessary for invariance under the 
scaling in \eqref{SCALEh} and \eqref{SCALEe}.  In this regard, I
observe that the eigenvalues of the Laplacian $\triangle_0$ themselves
scale with $\Lambda$ as 
\begin{equation}\label{EVALS}
\lambda \,\longmapsto\, \Lambda^{-1} \, \lambda\,.
\end{equation}

When ${\phi:\BR\times\Sigma \to S^1}$ has non-trivial winding, the
classical mode expansion in \eqref{EIGEN} must be generalized only
slightly.  Exactly as in Section $2.2$ in \cite{BeasleyI:2014}, we
consider a harmonic representative for the cohomology class ${\omega
  \in H^1(\Sigma;\BZ)}$.  Mildly abusing notation, I re-use
$\omega$ to refer to this representative in the space $\CH^1(\Sigma)$
of harmonic one-forms on $\Sigma$.  Associated to the harmonic one-form $\omega$
with integral periods on $\Sigma$ is a fiducial harmonic map
${\Phi_\omega:\Sigma\to S^1}$ satisfying 
\begin{equation}\label{HARM1}
\frac{d\Phi_\omega}{2\pi} \,=\, \omega\,,\qquad\qquad \omega
\,\in\,\CH^1(\Sigma)\,.
\end{equation}
Since ${d^\dagger\omega = 0}$, we see that ${\triangle_0\Phi_\omega =
  d^\dagger d\Phi_\omega \,=\, 0}$ automatically.
By Hodge theory, the map $\Phi_\omega$ is determined
by $\omega$ up to a constant.  To fix that constant, we select a
basepoint ${\sigma_0\in\Sigma}$, which will re-occur later in Section
\ref{Maxwell}, and we impose 
\begin{equation}\label{HARM2}
\Phi_\omega(\sigma_0) \,=\, 0\,\,\mod\, 2\pi\,.
\end{equation}
Together, the conditions in \eqref{HARM1} and \eqref{HARM2} uniquely determine
the fiducial map $\Phi_\omega$ with given winding.

Because the winding-number is additive, the general solution to
the equation of motion in \eqref{EOFMP} with winding-number
${\omega\in\BL}$ can now be written as the sum of the
topologically-trivial solution in \eqref{EIGEN} with the fiducial
harmonic map $\Phi_\omega(z,\bar z)$,
\begin{equation}\label{EIGENOM}
\phi(t,z,\bar z) \,=\, \Phi_\omega(z,\bar z) \,+\, \frac{\phi_0}{e^2 \ell} \,+\,
\frac{2\pi t}{\ell}\,p_0 \,+\, \sqrt{\pi}\,\sum_{\lambda > 0}
\frac{e^2}{\lambda}\,\psi_\lambda(z,\bar z) 
\,\Big[a_\lambda \, \e{-i\,\lambda\,t} \,+\, a_\lambda{}^{\!\dagger} \,
\e{i\,\lambda\,t}\Big].
\end{equation}
In these terms, the coefficients $\phi_0$, $p_0$, and
$(a_\lambda,\,a_\lambda{}^{\!\dagger})$ for all ${\lambda >
  0}$ parametrize the classical phase space for maps ${\phi:\BR \to
  \CX_\omega}$.

\medskip\noindent{\sl Sigma Model Hilbert Space}\smallskip

To quantize, we promote both the scalar field $\phi$ and
the momentum $\Pi_\phi$ in \eqref{MOMENTUM} to operators which obey the
canonical commutation relations 
\begin{equation}\label{COMPHI}
\big[\phi(z),\,\Pi_\phi(w)\big] \,=\, i \,
\delta_\Sigma(z, w)\,,\qquad\qquad z,w \,\in\,\Sigma\,.
\end{equation}
Here $\delta_\Sigma$ is a delta-function with support on the diagonal
${\Delta \subset \Sigma \times \Sigma}$.  We will take either of two
perspectives on \eqref{COMPHI}.  

From the first perspective, the commutator in \eqref{COMPHI} can be realized through the 
functional identification 
\begin{equation}
\Pi_\phi(w) \,=\, -i \, \frac{\delta}{\delta\phi(w)}\,,
\end{equation}
or equivalently via \eqref{MOMENTUM},
\begin{equation}\label{dPhidt}
\frac{e^2}{2\pi} \, \partial_t\phi(w) \,=\, -i \, \frac{D}{D\phi(w)}\,.
\end{equation}
Here $D/D\phi(w)$ is interpreted as a covariant functional
derivative incorporating the shift by $\theta$ in the canonical momentum,
\begin{equation}\label{DOVDPHI}
\frac{D}{D\phi(w)} \,=\, \frac{\delta}{\delta\phi(w)} \,-\, i\,
\frac{\theta}{2\pi \ell^2}\,.
\end{equation}
Because $\theta$ is just a constant,
\begin{equation}
\left[\frac{D}{D\phi(z)},\,\frac{D}{D\phi(w)}\right] =\,
0\,,\qquad\qquad z\,\neq\,w\,.
\end{equation}
As we will see quite explicitly, $D/D\phi$ thus describes a flat connection with non-trivial
holonomy over the configuration space $\CX$ of maps from $\Sigma$ to $S^1$.

For the alternative perspective on the commutator in \eqref{COMPHI}, we rewrite the
delta-function $\delta_\Sigma$ in terms of the orthonormal eigenbasis 
$\left\{\psi_\lambda\right\}$ for the scalar Laplacian,
\begin{equation}\label{HEATK}
\delta_\Sigma(z,w) \,=\, \frac{1}{\ell^2} \,+\, \sum_{\lambda > 0} e^4
\, \psi_\lambda(z) \, \psi_\lambda(w)\,,\qquad\qquad z,w\,\in\,\Sigma\,.
\end{equation}
The first term on the right in \eqref{HEATK} arises from the
constant mode $\psi_0$, and the factor of $e^4$ in the sum over the
higher eigenmodes is a result of the normalization condition in
\eqref{NORM}.  

After substituting the mode expansions in \eqref{EIGENOM} and
\eqref{HEATK} into the canonical commutation relation, we find that
$\left(\phi_0,p_0\right)$ and $(a_\lambda,a_\lambda{}^{\!\dagger})$ for ${\lambda > 0}$
satisfy the free-field Heisenberg algebra
\begin{equation}\label{HEISI}
\begin{aligned}
&\left[\phi_0,\,p_0\right] \,=\, i\,,\\
&\left[a_\lambda,\,a_{\lambda'}{}^{\!\dagger}\right] \,=\,
\frac{\lambda}{e^2} \, \delta_{\lambda \lambda'}\,,
\end{aligned}
\end{equation}
with all other commutators vanishing identically.  As usual, in the
second line of \eqref{HEISI} we introduce the Kronecker delta, defined by 
${\delta_{\lambda \lambda'} = 1}$ if ${\lambda = \lambda'}$ and 
${\delta_{\lambda \lambda'} = 0}$ otherwise. 

These commutation relations hold in each
winding-sector, independent of the class ${\omega \in 
  H^1(\Sigma;\BZ)}$, so the quantization will also be
independent of $\omega$.  As the counterpart to the topological
decomposition of ${\CX = \Map(\Sigma,S^1)}$ in \eqref{COMPONENT}, the
total Hilbert space $\SH_\Sigma$ for the periodic scalar field on
$\Sigma$ decomposes into the direct sum 
\begin{equation}\label{DIRSM}
\SH_\Sigma \,=\, \bigoplus_{\omega\in\BL} \,
\SH_\Sigma^{\,\omega}\,,\qquad\qquad \BL = H^1(\Sigma;\BZ)\,,
\end{equation}
where each subspace $\SH_\Sigma^{\,\omega}$ is itself a tensor product
(independent of $\omega$)
\begin{equation}\label{TENSORP}
\SH_\Sigma^{\,\omega} \,=\, {\mathsf H}_0 \otimes \bigotimes_{\lambda
  > 0} {\mathsf H}_\lambda\,.
\end{equation}
Of the two factors in the tensor product, ${\mathsf H}_\lambda$ is the
less interesting.  Up to an irrelevant choice of 
normalization, $a_\lambda$ and $a_\lambda{}^{\dagger}$ in \eqref{HEISI} satisfy the usual
commutator algebra for a harmonic oscillator with frequency
$\lambda$.  Hence ${\mathsf H}_\lambda$ is the Fock space for that
oscillator.

More interesting for us is the universal factor ${\mathsf H}_0$.  This
factor arises from the quantization of the zero-modes 
$\left(\phi_0,p_0\right)$ of the scalar field and thus does not depend upon the spectral
geometry of the surface $\Sigma$.  Together $\phi_0$ and $p_0$
simply describe the position and momentum of a free particle moving on
a circle with radius ${e^2 \ell}$.  The corresponding phase space is
the cotangent bundle $T^*S^1$ with the canonical symplectic structure,
and at least when ${\theta = 0}$ in \eqref{TOPI}, the quantization is
entirely standard.  Directly, 
\begin{equation}\label{LTWOS}
{\mathsf H}_0 \,\simeq\, L^2(S^1;\BC)\,,\qquad\qquad [\,\theta
= 0\,]
\end{equation}
spanned by the Fourier wavefunctions 
\begin{equation}\label{FOURIER}
\Psi_m(\phi_0) \,=\,
\exp{\!\left(i\,\frac{m}{e^2\ell}\,\phi_0\right)}\,,\qquad\qquad m
\,\in\,\BZ\,.
\end{equation}
As usual, the classical momentum $p_0$ becomes identified with the operator
${-i\,\partial/\partial\phi_0}$.  Via the identification
${{\bf P} = e^2 \ell \,p_0}$ in \eqref{BIGPp}, each Fourier wavefunction in
\eqref{FOURIER} is an eigenstate of the total momentum operator
\begin{equation}
{\bf P} \,=\, -i\, e^2 \ell \,\, \frac{\partial}{\partial\phi_0}\,.
\end{equation}

When the topological parameter $\theta$ is non-zero,
the quantization of $\phi_0$ and $p_0$ is modified.  After we project 
\eqref{dPhidt} and \eqref{DOVDPHI} to the space of zero-modes, $p_0$
becomes identified with the $\theta$-dependent operator 
\begin{equation}\label{DDTHETA}
p_0 \,=\, -i\,\frac{D}{D\phi_0}\,,\qquad\qquad\qquad \frac{D}{D\phi_0} \,=\,
\frac{\partial}{\partial\phi_0} \,-\, i\,\frac{\theta}{2\pi e^2 \ell}\,.
\end{equation}
Evidently, $D/D\phi_0$ in \eqref{DDTHETA} is the covariant derivative
for a unitary flat connection on a complex line-bundle $\CL$ over the
circle, with holonomy
\begin{equation}\label{HOLSONE}
\Hol_{S^1}\!\left({D}/{D\phi_0}\right) =\, \exp{\!\left(i\,\theta\right)}\,.
\end{equation}
As the natural generalization of \eqref{LTWOS}, the zero-mode Hilbert space ${\mathsf
  H}_0$ is the space of square-integrable sections of $\CL$,
\begin{equation}\label{HZEROL}
{\mathsf H}_0 \,=\, L^2(S^1;\CL)\,,
\end{equation}
on which ${\bf P}$ now acts covariantly by 
\begin{equation}\label{BFPOP}
{\bf P} \,=\, -i\,e^2 \ell\,\frac{D}{D\phi_0}\,.
\end{equation}
For the Fourier wavefunction $\Psi_m(\phi_0)$ in \eqref{FOURIER}, all this is
just to say that 
\begin{equation}\label{BFPOPII}
{\bf P}\cdot\Psi_m(\phi_0) \,=\, \left(m \,-\,
  \frac{\theta}{2\pi}\right)\cdot\Psi_m(\phi_0),\qquad\qquad m \,\in\,\BZ\,,
\end{equation}
as follows directly from \eqref{DDTHETA}.  Hence the topological
parameter $\theta$ induces a uniform shift on the eigenvalues of ${\bf
  P}$ away from integral values.  Manifestly, the spectrum of ${\bf
  P}$ depends only on the value of $\theta$ modulo $2\pi$.

Because the zero-mode Hilbert space ${\mathsf H}_0$ is graded by ${\bf
  P}$, the full sigma model Hilbert space $\SH_\Sigma$ is
bigraded by the lattice ${\BZ \oplus \BL}$,
\begin{equation}\label{LathH}
\SH_\Sigma \,\simeq \bigoplus_{(m,\,\omega)\in\BZ\oplus\BL}
\SH_\Sigma^{\,m, \omega}\,,
\end{equation}
in parallel to the bigrading by ${\BZ\oplus\BZ}$ for
$\SH_{S^1}$ in \eqref{DSUM}.  As a convenient shorthand, I let
$|m;\omega\rangle$ denote the Fourier wavefunction $\Psi_m(\phi_0)$,
considered in the topological sector with winding-number $\omega$ and
satisfying the vacuum condition
\begin{equation}\label{FockVcs}
a_\lambda|m;\omega\rangle = 0\,,\qquad\qquad \lambda
  > 0\,.
\end{equation}
All other Fock states in $\SH_\Sigma$ are obtained by acting with the
oscillator raising-operators $a_\lambda{}^{\!\dagger}$ on each Fock vacuum
$|m;\omega\rangle$, so the summands in $\SH_\Sigma$ above are more explicitly
\begin{equation}\label{CURLYHS}
\SH_\Sigma^{\,m,\omega} \,=\, \BC\cdot|m;\omega\rangle \otimes \bigotimes_{\lambda > 0}
{\mathsf H}_\lambda\,.  
\end{equation}

Philosophically, the grading by the eigenvalue ${m}$ in \eqref{LathH} is a quantum grading (since
we must quantize $\phi_0$ to define it!), whereas the grading by the
winding-number ${\omega}$ is classical, just as we saw in Section \ref{Intro}
for quantization on $S^1$.  But needless to say, because ${\BZ
  \neq \BL \simeq \BZ^{2g}}$, duality on $\Sigma$ cannot
exchange the two gradings, as occurs for duality on $S^1$.  Rather,
the role of duality will be to exchange the quantum versus classical
interpretations of each.

Finally, let us consider the action of the sigma model Hamiltonian ${\bf H}$ on
the Hilbert space.  After the identification in \eqref{dPhidt}, the
classical Hamiltonian becomes the operator
\begin{equation}
{\bf H} \,=\, \int_\Sigma
\left[-\frac{\pi}{e^2}\,\frac{D^2}{D\phi^2}\,\vol_\Sigma \,+\,
  \frac{e^2}{4\pi}\,d\phi\^\*d\phi \,-\,
  \frac{e^2}{2\pi}\,\upalpha\^d\phi\right].
\end{equation}
Upon substituting for the momentum ${\bf P}$ in \eqref{BFPOP},
\begin{equation}\label{QMHII}
{\bf H} \,=\, \int_\Sigma\left[\frac{\pi}{e^2} \left(\frac{{\bf P}^2}{\ell^4} +
    \cdots\right) \vol_\Sigma
  \,+\, \frac{e^2}{4\pi}\,d\phi\^\*d\phi \,-\,
  \frac{e^2}{2\pi}\,\upalpha\^d\phi \right],
\end{equation}
where the ellipses indicate terms in $D^2/D\phi^2$ which involve the
non-zero eigenmodes of $\phi$ and thus the Fock operators
$(a_\lambda,a_\lambda{}^{\!\dagger})$.  

The spectrum of ${\bf H}$ depends upon the corresponding spectrum of
eigenvalues $\{\lambda^2\}$ for the scalar Laplacian
$\triangle_0$, which in turn depends upon the geometry of $\Sigma$.
To simplify the situation, we consider the action of ${\bf H}$ only
on the Fock vacua $|m;\omega\rangle$ in \eqref{FockVcs}.  From
\eqref{BFPOPII} and \eqref{QMHII},
\begin{equation}\label{SPECTH}
{\bf H}\,|m;\omega\rangle \,=\,
e^2\left[\frac{\pi}{(e^2\ell)^2}\left(m - \frac{\theta}{2\pi}\right)^2 +\,
  \pi\left(\omega,\omega\right) \,-\, 
  \left\langle\upalpha,\omega\right\rangle +\,
  \frac{E_0}{e^2\ell}\right]\!|m;\omega\rangle\,.
\end{equation}
Here $\left(\omega,\omega\right)$ is the $L^2$-norm of the harmonic one-form
  appearing in \eqref{HARM1},
\begin{equation}\label{L2OM1}
\left(\omega,\omega\right) =\, \int_\Sigma
\omega\^\*\omega\,,\qquad\qquad \omega\,\in\,\CH^1(\Sigma)\,,
\end{equation}
and $\langle\upalpha,\omega\rangle$ denotes the intersection pairing
\begin{equation}\label{INTERSECT}
\left\langle\upalpha, \omega\right\rangle \,=\, \int_\Sigma
\upalpha\^\omega\,,\qquad\qquad \upalpha \,\in\,\CH^1(\Sigma)\,.
\end{equation}
As will be important later, note that $\left(\omega,\omega\right)$ in
\eqref{L2OM1} is a conformal invariant, for which only the complex structure
on $\Sigma$ matters, and of course $\left\langle\upalpha, \omega\right\rangle$
is purely topological.

The energy in \eqref{SPECTH} also includes a constant term $E_0$,
independent of $e^2$, $m$, and $\omega$, which arises from the sum over the zero-point
energies $\ha\lambda$ of each oscillating 
eigenmode of $\phi$.  The factor of $1/\ell$ which multiplies $E_0$
is fixed by the scaling in \eqref{EVALS}, and we have pulled out an overall factor of $e^2$ from
${\bf H}$ so that the quantity in brackets is scale-invariant (or dimensionless).
Physically, $E_0/\ell$ is a Casimir energy on the
compact surface $\Sigma$, and some method of regularization must be chosen to
make sense of the divergent sum ${E_0 \sim \sum_{\lambda > 0} \ha 
  \lambda}$ which naively defines 
$E_0$, eg.~by normal-ordering or use of the zeta-function.
For comparison under duality, the particular method used to define
$E_0$ will not matter, so we simply assume that $E_0$ has
been determined in some way from the non-zero eigenvalues of the
scalar Laplacian $\triangle_0$ on $\Sigma$.

Finally, let us consider the dependence of the spectrum of ${\bf H}$
on the effective coupling ${1/e^2 \ell}$.  Though the
abelian sigma model is a free quantum field theory, there remains a
definite sense in which the spectrum simplifies in the
weakly-coupled regime ${1/e^2\ell \ll 1}$.  As apparent
from \eqref{SPECTH}, in this limit the quantum states with least
energy in any given topological sector are precisely the Fock vacua
$|m;\omega\rangle$, for arbitrary values of the Fourier momentum ${m
  \in \BZ}$.  

Conversely, when $1/e^2\ell$ is of order-one, we do not find a clean
separation in energy between the Fock vacua $|m;\omega\rangle$ and oscillator
states such as $a_\lambda{}^{\!\dagger}|0;\omega\rangle$ for suitable
$\lambda$.  Hence in the latter case, the low-lying energy spectrum of
the quantum sigma model depends much more delicately on the geometry
of $\Sigma$.

\subsection{The Quantum Maxwell Theory}\label{Maxwell}

We now consider the quantization of Maxwell theory on ${M =
  \BR \times \Sigma}$, with the same Lorentzian product metric already
appearing in \eqref{3DM}.

Classically, the Maxwell gauge field $A$ is a connection on a fixed
principal $U(1)$-bundle $P$ over $M$, 
\begin{equation}
\begin{aligned}
U(1) \rightarrow\,\,\, &P\\[-8 pt]
&\mskip -5mu \downarrow\\[-1 ex]
&\mskip -6mu M
\end{aligned}\quad.
\end{equation}
When ${M = \BR \times \Sigma}$, the restriction of $P$ determines an
associated complex line-bundle $L$ over $\Sigma$, with Chern class
\begin{equation}
c_1(L) \,=\, \left[\frac{F_A}{2\pi}\right]\,\in\,
H^2(\Sigma;\BZ)\,.
\end{equation}
Here ${F_A = dA}$ is the curvature, and we specify the Chern class of
$L$ by a single integer
\begin{equation}\label{DEGREE}
m \,=\, \deg(L)\,\in\,\BZ\,.
\end{equation}
The coincidence in notation between $m$ in \eqref{FOURIER} and
\eqref{DEGREE} is no accident.

The integer $m$ suffices to fix the topological type of both $P$ and
$L$.  However, for purpose of quantization, we will need to endow the
line-bundle $L$ with a holomorphic structure as well.  Because
$\Sigma$ carries a complex structure associated to its Riemannian
metric $h$ as in 
\eqref{HODGEII}, $L$ can be given a holomorphic structure uniformly
for all degrees as soon as we pick a basepoint ${\sigma_0\in\Sigma}$.
We set
\begin{equation}
L \,=\, \CO_\Sigma(m\,\sigma_0)\,,\qquad\qquad \sigma_0 \,\in\,\Sigma\,.
\end{equation}
By definition, holomorphic sections of $L$ can be identified with meromorphic functions
on $\Sigma$ which have a pole of maximum degree $m$ at the point
${\sigma_0\in\Sigma}$.  

Note that the choice of basepoint is only
relevant when $\Sigma$ has genus ${g \ge 1}$, since the holomorphic
structure on any line-bundle of degree $m$ over $\BC\BP^1$ is unique.
The same remark also applies to our previous choice of basepoint for the sigma
model: in genus zero, the only fiducial harmonic map $\Phi_\omega$ is
constant, so the condition in \eqref{HARM2} does not actually depend
upon the choice of $\sigma_0$.

Specialized to ${M = \BR \times \Sigma}$, the free Maxwell action becomes
\begin{equation}\label{MAXWELL}
{\bf I}_0(A) \,=\, \frac{1}{4\pi
  e^2}\,\int_{\BR\times\Sigma}\!\!dt\,\big[E_A\^\*E_A \,-\,
F_A\^\*F_A\big]\,,\qquad\quad E_A\,=\, \iota_{\partial/\partial
  t}F_A \,\in\, \Omega^1_\Sigma\,.
\end{equation}
Here we stick to the assumption that ${\*\equiv \*_\Sigma}$ is the
Hodge operator on $\Sigma$, so we have separated the curvature into the electric
component $E_A$, which transforms like a one-form on $\Sigma$, along
with the magnetic component ${F_A \equiv F_A|_\Sigma}$, which transforms 
like a two-form on $\Sigma$.  Explicitly in local coordinates,
\begin{equation}
E_A \,=\, F_{A,t z} \, dz
  \,+\, F_{A,t \bar z} \,\, d\bar z\,.
\end{equation}
Invariance under the scaling in \eqref{SCALEh}, \eqref{SCALEe}, and
\eqref{SCALEt} fixes the dependence of the Maxwell action on the
electric coupling $e^2$, and the overall factor of $1/4\pi$ in
\eqref{MAXWELL} appears by convention.

As in Section \ref{Scalar}, topological terms can also be added to the
Maxwell action, of the form 
\begin{equation}\label{TOPMAX}
{\bf I}_1(A) \,=\, \frac{1}{2\pi} \int_{\BR \times
  \Sigma}\!\!dt\,\upbeta\^E_A \,+\, \frac{\theta}{2\pi e^2 \ell^2}
\int_{\BR\times\Sigma}\!\!dt\, F_A\,.
\end{equation}
Like $\upalpha$ in \eqref{ALPHP}, $\upbeta$ is a real harmonic
one-form,
\begin{equation}
\upbeta \,\in\, \CH^1(\Sigma)\,,
\end{equation}
and ${\theta\in\BR}$ is a real parameter that will correspond under
duality to the angle already appearing in \eqref{SPECTH}.  With some
malice aforethought, the coefficient $1/e^2\ell^2$ in \eqref{TOPMAX}
has been chosen to achieve this identification, along with invariance
under the scaling in \eqref{SCALEh}, \eqref{SCALEe}, and \eqref{SCALEt}.

We then consider the total gauge theory action 
\begin{equation}
{\bf I}_{\rm tot}(A) \,=\, {\bf I}_0(A) \,+\, {\bf I}_1(A)\,,
\end{equation}
or more explicitly,
\begin{equation}\label{MAXWELI}
{\bf I}_{\rm tot}(A) \,=\, \frac{1}{4\pi
  e^2}\int_{\BR\times\Sigma}\!\!dt\, \left[E_A\^\*E_A \,-\, F_A\^\*F_A
\,-\, 2 e^2 E_A\^\upbeta \,+\, 2 \frac{\theta}{\ell^2} F_A\right]\,.
\end{equation}
Previously for the periodic scalar field, the
angular parameter $\theta$ served to modify the definition
\eqref{MOMENTUM} of the canonical momentum $\Pi_\phi$.  This role is
now taken by the harmonic one-form $\upbeta$, which appears in the
canonical momentum 
\begin{equation}\label{MOMENTA}
\Pi_A \,=\, \frac{1}{2\pi e^2} \, \*E_A -
\frac{1}{2\pi}\,\upbeta\,.
\end{equation}
In terms of $\Pi_A$, the classical Hamiltonian is\footnote{The
  superscript on ${\bf H}^\vee$ in \eqref{BIGHA} serves to differentiate
  the Maxwell Hamiltonian notationally from the Hamiltonian ${\bf H}$ for the
periodic scalar field in Section \ref{Scalar}.} 
\begin{equation}\label{BIGHA}
{\bf H}^\vee \,=\, \int_\Sigma \left[\pi e^2\left(\Pi_A \,+\,
    \frac{\upbeta}{2\pi}\right)\!\^\*\!\left(\Pi_A \,+\,
    \frac{\upbeta}{2\pi}\right) + \frac{1}{4\pi e^2} \,F_A\^\*F_A
  \,-\, \frac{\theta}{2\pi e^2 \ell^2}\,F_A\right].
\end{equation}
The degree $m$ of the line-bundle $L$ is measured by the net magnetic flux
through the surface, 
\begin{equation}
\int_\Sigma \frac{F_A}{2\pi} \,=\, m\,,
\end{equation}
so $\theta$ in \eqref{BIGHA} now serves to distinguish the
topological sectors labelled by $m$.

Because we have yet to fix a gauge, the classical Maxwell Hamiltonian ${\bf
  H}^\vee$ is degenerate along gauge orbits.  As a
remedy, we work throughout in Coulomb gauge, 
\begin{equation}\label{COULOMB}
A_t \,=\, 0\,,
\end{equation}
where $A_t$ is the time-component of the gauge field on ${M =
  \BR \times \Sigma}$.  In Coulomb gauge, the equation of motion for
$A_t$ holds identically as the Gauss law constraint
\begin{equation}\label{GAUSS}
d{}^{\dagger}E_A \,=\, 0\,.
\end{equation}
(Because ${d\upbeta=0}$, the topological terms do not modify the Gauss 
law on $\Sigma$.)  On $\BR^{1,2}$, Coulomb gauge does not respect Lorentz
invariance, which is the main disadvantage of Coulomb
gauge.  For quantization on ${M = \BR \times \Sigma}$, though,
Lorentz invariance is neither here nor there, and the gauge condition in
\eqref{COULOMB} is perfectly natural.

To fix the remaining time-independent gauge transformations on
$\Sigma$, we impose the further harmonic condition 
\begin{equation}\label{HARMO}
d{}^{\dagger}A \,=\, 0\,.
\end{equation}
Harmonic gauge on $\Sigma$ is particularly convenient from the geometric
perspective.  In this gauge, the Gauss constraint in \eqref{GAUSS} is
automatically obeyed, and $A$ satisfies the classical wave equation 
\begin{equation}\label{WAVEA}
\partial{}_t^2 A \,+\, \triangle_1 A \,=\, 0\,,\qquad\qquad A
\,\in\,\Omega^1_\Sigma\,,
\end{equation}
where ${\triangle_1 = d{}^{\dagger} d + d d{}^{\dagger}}$ is the de Rham
Laplacian for one-forms on $\Sigma$.  We considered precisely 
the same equation of motion in \eqref{EOFMP} for the periodic scalar
field $\phi$, so quantization of $A$ in
harmonic gauge will share many features with quantization of
$\phi$, and duality will be manifest.

Finally, if $A$ is any time-independent connection on $\Sigma$, the
equation of motion in \eqref{WAVEA} implies that the curvature 
is also harmonic,
\begin{equation}\label{HARMF}
d{}^{\dagger}F_A \,=\, 0\,,\qquad\qquad F_A \,\in\,\Omega^2_\Sigma\,.
\end{equation}
Thus the classical vacua of Maxwell theory on $\Sigma$ correspond to
harmonic connections on the line-bundle $L$.

When $\Sigma$ has genus ${g \ge 1}$, Maxwell theory on ${M = \BR
  \times \Sigma}$ is invariant under a continuous $U(1)^{2 g}$ global symmetry,
which does not occur at genus zero.  To describe the action of the
symmetry on the gauge field, we first select an integral harmonic
basis $\left\{\Fe_1,\ldots,\Fe_{2g}\right\}$ for the cohomology
lattice
\begin{equation}\label{HARMEs}
\BL \,\simeq\, \BZ \Fe_1 \oplus \cdots \oplus \BZ \Fe_{2
  g}\,,\qquad\qquad \BL = H^1(\Sigma;\BZ)\,.
\end{equation}
The group $U(1)^{2g}$ then acts on the gauge field by shifts
\begin{equation}\label{JACS}
U(1)^{2 g}:\quad A \,\longmapsto\, A \,+\, \sum_{j=1}^{2g} c_j\,
\Fe_j\,,\qquad\qquad c_j \,\in\,\BR/2\pi\BZ\,.
\end{equation}
Such shifts for any constant $c_j$ trivially preserve both the Maxwell
action in \eqref{MAXWELI} and the gauge conditions in \eqref{COULOMB}
and \eqref{HARMO}.  Note also that shifts by elements in
the lattice $2\pi\BL$ are induced by homotopically non-trivial,
``large'' gauge transformations 
\begin{equation}
A^u \,=\, A \,+\, i\, u^{-1} \, du\,,\qquad\qquad
u:\Sigma\,\to\,U(1)\,,
\end{equation}
so these lattice elements act as the identity modulo gauge-equivalence.
As a result, the parameters $c_j$ in \eqref{JACS} are circle-valued,
and the global symmetry group is compact.  

The group $U(1)^{2g}$ acts to shift the holonomies of the gauge field, so we
can think of this global symmetry group more intrinsically as the
Jacobian torus of $\Sigma$,
\begin{equation}\label{JACOBIAN}
\SJ_\Sigma \,=\, H^1(\Sigma;\BR)/2\pi\BL\,\simeq\, U(1)^{2
  g},\qquad\qquad \BL \,=\, H^1(\Sigma;\BZ)\,.
\end{equation}
The Jacobian $\SJ_\Sigma$, in its role as the moduli space of flat
$U(1)$-connections on $\Sigma$, will be essential in
analyzing abelian duality at higher genus.

The $U(1)^{2 g}$ global symmetry of Maxwell theory on $\Sigma$ is the
counterpart to the more obvious $U(1)$ symmetry of the periodic
scalar field in \eqref{UONEL}.  Just as for the conserved momentum
${\bf P}$ in \eqref{BIGP}, the global symmetry of Maxwell
theory leads to a set of $2g$ conserved charges 
\begin{equation}\label{BIGDULP}
\begin{aligned}
{\bf W_j} \,&=\, \frac{1}{2\pi e^2} \int_\Sigma \Fe_j
\^\*E_A\,,\qquad\qquad\qquad {\bf j}\,=\,1\,,\ldots,2g\,,\\
&=\, \int_\Sigma \Fe_j\^\!\left(\Pi_A \,+\, \frac{\upbeta}{2\pi}\right).
\end{aligned}
\end{equation}
Conservation of ${\bf W_j}$ follows from the harmonic condition 
${d\Fe_j = d{}^{\dagger}\Fe_j = 0}$, as well as the classical equation of motion
in \eqref{WAVEA}.  

Because of the $U(1)^{2g}$ global symmetry, upon
quantization the Maxwell Hilbert space will automatically carry an
integral grading by the eigenvalues of ${\bf W_j}$.  Moreover,
since $\*E_A$ is directly related \eqref{MOMENTA} to the canonical
momentum $\Pi_A$ for the gauge field, the grading by ${\bf
  W_j}$ will again be interpreted physically as a quantum
grading by total momentum.

\medskip\noindent{\sl Classical Mode Expansion}\medskip

This background material out of the way, we now quantize the Maxwell gauge
field on the Riemann surface $\Sigma$.  As for the periodic scalar
field, our main interest lies in a universal set of low-lying
energy levels which are not sensitive to the detailed spectral
geometry of $\Sigma$.

The quantization of $A$ on $\Sigma$ involves a countable number of
topological sectors, labelled by the degree $m$ of the line-bundle
$L$.  By analogy to the decomposition \eqref{COMPONENT} of the
configuration space for the scalar field, we write the configuration
space for the Maxwell gauge field as a union of components 
\begin{equation}
\CA \,=\, \bigsqcup_{m\in\BZ} \CA_m\,,
\end{equation}
where $\CA_m$ is the affine space of unitary connections on the
complex line-bundle $L$ of degree $m$ over $\Sigma$,
\begin{equation}\label{BIGAM}
\CA_m \,=\, \left\{ A \,\in\,\CA \,\Biggr|\, \int_\Sigma F_A \,=\,
  2\pi m\right\}\,.
\end{equation}
Each connected component ${\CA_m\subset \CA}$ of the configuration space
must be quantized separately, so the Maxwell Hilbert space
$\SH^\vee_\Sigma$ automatically carries an integral grading by the
degree ${m\in\BZ}$.

Having broken our quantization problem into countably-many
pieces, we solve the classical equation of motion \eqref{WAVEA}
for $A$ in harmonic gauge.  As a special case, we begin by considering only
time-independent classical solutions, corresponding to connections
with harmonic curvature on $\Sigma$.  

When $L$ has degree ${m = 0}$ and hence is topologically trivial, a
harmonic connection on $L$ is simply a flat connection, of the form 
\begin{equation}\label{HARMA}
A \,=\, \sum_{j=1}^{2 g}\, \varphi_0{}^j \, \Fe_j\,,\qquad\qquad \varphi_0^j \,\in\, \BR/2\pi\BZ\,,
\end{equation}
for the fixed harmonic basis  $\left\{\Fe_1,\cdots,\Fe_{2 g}\right\}$
of ${H^1(\Sigma;\BZ)}$.  The expansion coefficients $\varphi_0{}^j$
for ${j=1,\ldots,2g}$ are thus angular coordinates on the Jacobian $\SJ_\Sigma$, 
which has already appeared in \eqref{JACOBIAN}.  Equivalently,
$\left(\varphi_0{}^1,\ldots,\varphi_0{}^{2g}\right)$ characterize the
holonomies of $A$ around a generating set of closed one-cycles on $\Sigma$.

Because $\Sigma$ carries a complex structure, each harmonic one-form
$\Fe_j$ in \eqref{HARMA} can be decomposed according to its
holomorphic/anti-holomorphic type via \eqref{HODGEII}, in which case
$\SJ_\Sigma$ itself inherits a complex structure.  Intrinsically as a
complex torus,
\begin{equation}\label{PICZ}
\begin{aligned}
\SJ_\Sigma \,&=\, H^1_{\bar\partial}(\Sigma,\CO_\Sigma)/2\pi\BL\,,\qquad\qquad
\BL \,=\, H^1(\Sigma;\BZ)\,,\\
&\simeq\, \Pic_0(\Sigma)\,,
\end{aligned}
\end{equation}
where $\Pic_0(\Sigma)$ denotes the group of isomorphism classes of
holomorphic line-bundles of degree zero on $\Sigma$, with group
multiplication given by the tensor product of line-bundles.

If $L$ has degree ${m \neq 0}$, then a harmonic connection on $L$ cannot be
flat.  Instead, the curvature $\widehat{F}{}_m$ of any harmonic connection on $L$
is proportional to the Riemannian volume form on $\Sigma$,
\begin{equation}\label{FHAT}
\widehat{F}{}_m \,=\, \frac{2\pi m}{\ell^2} \,
\vol_\Sigma\,,
\end{equation}
where the proportionality constant in \eqref{FHAT} is determined by
the topological condition in \eqref{BIGAM}. The formula in \eqref{FHAT} is
insufficient to fix a fiducial $U(1)$-connection $\widehat{A}{}_m$
with the given curvature, since $\widehat{A}{}_m$ may have non-trivial
holonomies not detected by $\widehat{F}{}_m$.  To fix
$\widehat{A}{}_m$ uniquely, we use our auxiliary choice of basepoint
${\sigma_0 \in \Sigma}$ and the resulting holomorphic identification
${L = \CO_\Sigma(m\,\sigma_0)}$.  Precisely the same choice appeared
in the quantization \eqref{HARM2} of the periodic scalar field, for
precisely the same reason.

Abstractly, the basepoint ${\sigma_0 \in \Sigma}$ provides an
isomorphism between distinct components of the Picard group of all
holomorphic line-bundles on $\Sigma$, 
\begin{equation}
\Pic(\Sigma) \,=\, \bigsqcup_{m\in\BZ} \Pic_m(\Sigma)\,,
\end{equation}
via the tensor product
\begin{equation}\label{ISOPIC}
\begin{aligned}
\otimes\CO_\Sigma(\sigma_0):\Pic_m(\Sigma)
\,&\buildrel\simeq\over\longrightarrow\,
\Pic_{m+1}(\Sigma)\,,\\
{\mathfrak L} \quad&\longmapsto\quad\!\! {\mathfrak L}\otimes
\CO_\Sigma(\sigma_0)\,.
\end{aligned}
\end{equation}
Here $\Pic_m(\Sigma)$ denotes the component of the Picard group
consisting of degree $m$ holomorphic line-bundles on $\Sigma$.  Under
the isomorphism in \eqref{ISOPIC}, all components of the Picard group
are identified with the distinguished component
${\Pic_0(\Sigma)\simeq\SJ_\Sigma}$.  Because we 
already have a fiducial connection in $\Pic_0(\Sigma)$, namely
${\widehat{A}{}_0 = 0}$ in \eqref{HARMA}, we just take
$\widehat{A}{}_m$ to be the image of $\widehat{A}{}_0$ under the
isomorphism.  Equivalently from the differential perspective,
$\widehat{A}{}_m$ is the unique harmonic, unitary connection
compatible with the holomorphic structure on
$\CO_\Sigma(m\,\sigma_0)$.  See for instance Ch.\,4 of
\cite{Friedman:98} for more about the existence and uniqueness of
$\widehat{A}_m$.

Our choice for the fiducial harmonic connection $\widehat{A}{}_m$ is
natural in the following sense.  Trivially, $\CO_\Sigma(m\,\sigma_0) =
\CO_\Sigma(\sigma_0)^{\otimes m}$.  Thus $\widehat{A}{}_m$ for general
$m$ is related to the basic connection $\widehat{A}{}_1$ on
$\CO_\Sigma(\sigma_0)$ by  
\begin{equation}
\widehat{A}_m \,=\, m \, \widehat{A}_1\,,\qquad\qquad m \,\in\,\BZ\,.
\end{equation}
This identity is clearly compatible with the formula for the harmonic
curvature $\widehat{F}{}_m$ in \eqref{FHAT}.  For the remainder of the
paper, I simplify the notation by setting ${\widehat{A} \equiv \widehat{A}{}_1}$.

According to the preceding discussion, in each degree ${m\in\BZ}$, the arbitrary
time-independent solution to the classical equation of motion for $A$ in
\eqref{WAVEA} is given up to gauge-equivalence by a point on the
Jacobian ${\SJ_\Sigma \simeq U(1)^{2 g}}$.  To describe the more
general, time-dependent solution in harmonic gauge, we perform an
expansion of $A$ in eigenmodes of the de Rham Laplacian $\triangle_1$,
\begin{equation}\label{EIGENA}
\begin{aligned}
A(t,z,\bar z) \,&=\, m\,\widehat{A} \,+\, \sum_{j=1}^{2g} \,
\varphi_0{}^j\,\Fe_j
\,+\, 2\pi e^2 \, t \sum_{j,k=1}^{2g} \, p_{0,j} \, \big(\SQ^{-1}\big){}^{j k} \,
\Fe_k \,\,+\, \\
&\qquad\,+\,\sqrt{\pi}\,\sum_{\lambda > 0} \frac{e^2}{\lambda}\,
\chi_\lambda(z,\bar z) \Big[a_\lambda \, \e{-i\,\lambda\,t} \,+\,
  a_\lambda{}^{\dagger} \, \e{i\,\lambda\,t}\Big].
\end{aligned}
\end{equation}
The eigenmode expansion of $A$ in \eqref{EIGENA} requires several comments.

First, ${p_{0,j}\in\BR}$ for ${j=1,\ldots,2g}$ are the classical momenta
conjugate to the angular coordinates $\varphi_0{}^j$ on the Jacobian.
The coefficient of $e^2$ which multiplies $p_{0,j}$ is fixed by
scaling, due to the explicit $t$-dependence.  Also, $\SQ$ is the
positive-definite, symmetric matrix of $L^2$
inner-products\footnote{The same notation for $\SQ$ and 
  $\SQ^{-1}$ is used in \cite{BeasleyI:2014}, but in reference to
  similar quantities defined on a closed three-manifold $M$.} 
\begin{equation}\label{BIGSQ}
\SQ_{j k} \,=\, \left(\Fe_j, \Fe_k\right) \,=\, \int_\Sigma
\Fe_j\^\*\Fe_k\,,\qquad\qquad j,k = 1,\ldots,2g\,.
\end{equation}
The inverse matrix $\SQ^{-1}$ satisfies ${\left(\SQ^{-1}\right){}^{\!i j}
  \,\SQ_{j k} = \delta^i_k}$.  Since ${E_A = \partial_t A}$ in Coulomb
gauge, the expansion of $A$ ensures that $p_{0,j}$ is equal to the
conserved charge ${\bf W_j}$ in \eqref{BIGDULP},
\begin{equation}
{\bf W_j} \,=\, p_{0,j}\,,\qquad\qquad j\,=\,1\,,\ldots,2g\,.
\end{equation}

Proceeding to the second line of \eqref{EIGENA},  $(a{}_\lambda,
a_\lambda{}^{\!\dagger})$ are a conjugate pair of complex parameters
associated to the oscillating modes of $A$, and the coefficient
${\sqrt{\pi} \, e^2/\lambda}$ has been chosen to simplify the
commutator algebra upon quantization.  We also introduce
an orthonormal basis of one-forms ${\chi_\lambda\in\Omega^1_\Sigma}$ which satisfy
the joint conditions 
\begin{equation}\label{HARMCH}
d{}^\dagger\chi_\lambda \,=\, 0\,,\qquad\qquad \chi_\lambda\,\in\,\Omega^1_\Sigma\,,
\end{equation}
as well as 
\begin{equation}\label{EIGENCH}
\triangle_1\chi_\lambda \,=\, \lambda^2 \, \chi_\lambda\,,\qquad\qquad
\lambda \,\in\,\BR\,,
\end{equation}
where by convention ${\lambda>0}$ is positive.  The first condition
instantiates the harmonic gauge in \eqref{HARMO}, and the second
condition states that $\chi_\lambda$ is an eigenform for the de Rham
Laplacian $\triangle_1$ acting on one-forms on $\Sigma$.

The eigenforms $\chi_\lambda$ have a simple relation to the
eigenfunctions ${\psi_\lambda\in\Omega^0_\Sigma}$ which appear in the
corresponding mode expansion for the scalar field $\phi$.  For if
$\psi_\lambda$ is an eigenfunction of the scalar Laplacian, 
\begin{equation}
\triangle_0\psi_\lambda \,=\, \lambda^2\,\psi_\lambda\,,\qquad\qquad
\lambda \,>\, 0\,,
\end{equation}
then we obtain a corresponding eigenform $\chi_\lambda$ for
$\triangle_1$ by setting 
\begin{equation}\label{HODGREL}
\chi_\lambda \,=\, \frac{e^2}{\lambda}\, \*d\psi_\lambda\,.
\end{equation} 
Since $d{}^{\dagger} = -\*d\*$ and ${\*^2=-1}$ on $\Omega^1_\Sigma$,
trivially ${d{}^{\dagger}\chi_\lambda = 0}$.  Also,
\begin{equation}\label{HODID}
\triangle_1\chi_\lambda \,=\, d{}^{\dagger} d\chi_\lambda \,=\,
-\*d\*d\left(\frac{e^2}{\lambda}\,\*d\psi_\lambda\right) \,=\,
\frac{e^2}{\lambda} \*d\left(\triangle_0\psi_\lambda\right) \,=\,
\lambda^2 \, \chi_\lambda\,.
\end{equation}
The relative coefficient ${e^2/\lambda}$ in \eqref{HODGREL} just ensures that
$\chi_\lambda$ has unit norm,
\begin{equation}
||\chi_\lambda||^2 \,=\, \int_\Sigma
\chi_\lambda\^\*\chi_\lambda \,=\, \frac{e^4}{\lambda^2} \int_\Sigma
\psi_\lambda \^\*\triangle_0\psi_\lambda \,=\, 1\,,
\end{equation}
assuming that $\psi_\lambda$ is normalized according to \eqref{NORM}.

Conversely, given any eigenform $\chi_\lambda$ satisfying
\eqref{HARMCH} and \eqref{EIGENCH} with ${\lambda> 0}$, we obtain
a normalized eigenfunction $\psi_\lambda$ via 
\begin{equation}\label{HODGRELII}
\psi_\lambda \,=\, -\frac{1}{e^2 \lambda} \, \*d\chi_\lambda\,.
\end{equation}
Following the same steps in \eqref{HODID}, one can verify directly
that $\psi_\lambda$ in \eqref{HODGRELII} is a normalized eigenfunction
of the scalar Laplacian $\triangle_0$.  The minus sign in
\eqref{HODGRELII} is a nicety which ensures that the map from 
$\psi_\lambda$ to $\chi_\lambda$ and back is the identity.

Together, the relations in \eqref{HODGREL} and \eqref{HODGRELII}
constitute a Hodge isomorphism between the non-zero spectrum of
$\triangle_0$ acting on $\Omega^0_\Sigma$ and the non-zero spectrum of
$\triangle_1$ acting on the intersection ${\Omega^1_\Sigma \cap
  \Ker(d{}^{\dagger})}$.  Consequently, the oscillator frequencies
${\lambda > 0}$ which appear in the harmonic expansion \eqref{EIGENA} of the
gauge field $A$ are precisely the same frequencies which appear in the
harmonic expansion \eqref{EIGEN} of the periodic scalar field
$\phi$.  This equality is essential for abelian duality to hold on $\Sigma$.

\medskip\noindent{\sl Canonical Commutation Relations}\medskip

Naively, to quantize the Maxwell theory, we promote both $A$ and the
canonical momentum $\Pi_A$ to operator-valued one-forms which satisfy
the canonical equal-time commutation relation 
\begin{equation}\label{CONAI}
\Big[A(z),\,\Pi_A(w)\Big] \,=\,
i\,\vol_\Sigma\cdot\delta_\Sigma(z,w)\,,\qquad\qquad z,w\,\in\,\Sigma\,.
\end{equation}
In making sense of \eqref{CONAI} geometrically, the Riemannian volume
form  on the right is to be interpreted as a section of the tensor product
${\vol_\Sigma\in\Omega^1_\Sigma\otimes\Omega^1_\Sigma}$, and
$\delta_\Sigma(z,w)$ remains the delta-function with support along the diagonal
${\Delta \subset \Sigma \times \Sigma}$.

Although convenient for our purposes, one feature of the canonical
commutator as written in \eqref{CONAI} is slightly
non-standard.  In local coordinates, the commutator would
typically be presented as ${[A_\mu(z),\,\wt\Pi_{A,\,\nu}(w)] = i \, h_{\mu \nu}
  \,\delta_\Sigma(z,w)}$ for ${\mu,\nu=1,2}$, where ${\wt\Pi_{A,\,\nu} =
  E_{A,\,\nu}/2\pi e^2}$ is the momentum conjugate to $A_\nu$ (assuming
${\upbeta=0}$ for simplicity), and $h_{\mu \nu}$ is the metric on
$\Sigma$.  In particular, this commutator is symmetric under the exchange
of the indices $\mu$ and $\nu$.  By contrast, the commutator in \eqref{CONAI} is
proportional not to the metric $h$ but to the volume form
${\vol_\Sigma \in \Omega^2_\Sigma}$ and hence is anti-symmetric under
the exchange of indices.  The difference in symmetry can be traced back to the
definition \eqref{MOMENTA} of ${\Pi_A = \*\wt\Pi_A}$, which involves
an extra factor of the Hodge star on $\Sigma$ relative to the standard
definition of the canonical momentum.  Though not entirely
conventional, our definition of $\Pi_A$ eliminates otherwise inelegant
factors of the Hodge star elsewhere.
  
In actuality, the situation is more complicated, because the
naive commutator in \eqref{CONAI} is not compatible with the Coulomb
gauge conditions 
\begin{equation}\label{CONAIA}
d{}^{\dagger}A \,=\, d{}^{\dagger}E_A \,=\, 0\,,\qquad\qquad A, E_A
\,\in\,\Omega^1_\Sigma\,,
\end{equation}
the latter of which implies 
\begin{equation}\label{CONAIPI}
d\Pi_A \,=\, d\!\left(\frac{1}{2\pi e^2} \, \*E_A -
\frac{1}{2\pi}\,\upbeta\right) =\, 0\,.
\end{equation}
I let $d_z$ and $d_w$ denote the respective de Rham operators acting
individually on the left and right factors in the product
$\Omega^1_\Sigma\otimes\Omega^1_\Sigma$.  Then the left-hand side of
\eqref{CONAI} is annihilated by $d_z{}^{\dagger}$ and $d_w$ according
to the gauge conditions in \eqref{CONAIA} and \eqref{CONAIPI}, but the
right-hand side is not.

This situation is a familiar feature of Coulomb gauge, as is the remedy.
Let $G(z,w)$ be the Green's function for the scalar Laplacian
$\triangle_0$ on $\Sigma$, such that
\begin{equation}\label{GREENS}
\triangle_0 G(z,w) \,=\, \delta_\Sigma(z,w) \,-\, \frac{1}{\ell^2}\,.
\end{equation}
Because $\Sigma$ is compact, we are careful to subtract the
contribution from the constant mode in \eqref{GREENS}, so
that both sides of the equation for $G(z,w)$ integrate to zero over
$\Sigma$.  Equivalently, $G(z,w)$ can be expanded in terms of the
orthonormal eigenbasis $\left\{\psi_\lambda\right\}$ for $\Omega^0_\Sigma$,
\begin{equation}\label{GZW}
G(z,w) \,=\, e^4 \sum_{\lambda > 0}\, \frac{\psi_\lambda(z) \,
  \psi_\lambda(w)}{\lambda^2}\,.
\end{equation}
For the ambitious reader who enjoys keeping track of factors of $e^2$,
recall that $e^2$ appears in the normalization condition \eqref{NORM}
for $\psi_\lambda$ and thus enters the expansion for $G(z,w)$.

The scalar Green's function $G(z,w)$ can be used to correct the naive
commutation relation in \eqref{CONAI} so that the right-hand side
is actually compatible with the Coulomb gauge constraints ${d{}^\dagger A =
d\Pi_A = 0}$.  To wit, the corrected commutator will be
\begin{equation}\label{CONAII}
\Big[A(z),\,\Pi_A(w)\Big] \,=\,
i\,\vol_\Sigma\cdot\delta_\Sigma(z,w) \,-\, i\left(d_z\!\otimes\!
  \*d_w\right)\!G(z,w)\,.
\end{equation}
Here $(d_z\!\otimes\!\*d_w)G(z,w)$ is the section of
${\Omega^1_\Sigma\otimes\Omega^1_\Sigma}$ obtained from the
action of the individual de Rham operators on $G(z,w)$.  In
terms of the spectral decomposition in \eqref{GZW},
\begin{equation}\label{SPECII}
\left(d_z\otimes \*d_w\right)\!G(z,w) \,=\, e^4
\sum_{\lambda > 0} \, \frac{d\psi_\lambda(z) \otimes
  \*d\psi_\lambda(w)}{\lambda^2} \,\in\, \Omega^1_\Sigma
\otimes \Omega^1_\Sigma\,.
\end{equation}
Most crucially, the right-hand side of the corrected commutator
\eqref{CONAII} does lie in the kernels of both $d_z{}^{\!\dagger}$ and $d_w$.
This statement can be verified either by direct computation from the
defining equation for $G(z,w)$ in \eqref{GREENS} or, as will be more
relevant here, by applying the spectral decomposition for
$(d_z\!\otimes\!\*d_w)G(z,w)$ in \eqref{SPECII}.  I take the latter approach.

In close analogy to \eqref{HEATK}, the term involving
the delta-function in \eqref{CONAII} can be presented as the sum
\begin{equation}\label{RESID}
\begin{aligned}
&\vol_\Sigma\cdot\delta_\Sigma(z,w) \,=\,\\ 
&\sum_{j,k=1}^{2g} 
(\SQ^{-1})^{j k} \,\Fe_j(z)\!\otimes\*\Fe_k(w) \,+\,
\sum_{\lambda > 0} \Big[\chi_\lambda(z)\otimes\*\chi_\lambda(w) \,-\,
  \*\chi_\lambda(z)\otimes\chi_\lambda(w)\Big]\,.
\end{aligned}
\end{equation}
Very briefly, the three terms on the right in \eqref{RESID} reflect
the three terms in the Hodge decomposition  
\begin{equation}\label{HODGD}
\Omega^1_\Sigma \,\simeq\, \CH^1(\Sigma) \oplus
\Im\!\big(d{}^{\dagger}|_{\Omega^2_\Sigma}\big) \oplus
\Im\!\big(d|_{\Omega^0_\Sigma}\big). 
\end{equation}
The first term in \eqref{RESID}, involving the harmonic forms $\Fe_j$
for ${j=1,\ldots,2g}$, describes the action of
${\vol_\Sigma\cdot\delta_\Sigma(z,w)}$ by wedge-product and
convolution on $\CH^1(\Sigma)$.  Otherwise,
the two sets of eigenforms $\left\{\chi_\lambda\right\}$ and
$\left\{\*\chi_\lambda\right\}$ for ${\lambda > 0}$ 
span the respective images of $d{}^{\dagger}$ and $d$, as
discussed previously in relation to \eqref{HODGREL}.  The latter two
terms on the right in \eqref{RESID} then account for the action of
${\vol_\Sigma\cdot\delta_\Sigma(z,w)}$ on
$\Im(d{}^{\dagger}|_{\Omega^2_\Sigma})$ and
$\Im(d|_{\Omega^0_\Sigma})$.  The relative minus sign in \eqref{RESID} can be
checked directly by working in a local frame on $\Sigma$ or
understood as a consequence of anti-symmetry under the exchange of the
factors in $\Omega^1_\Sigma\otimes\Omega^1_\Sigma$.

Because $\Fe_j$ is harmonic and ${d{}^{\dagger}\chi_\lambda =
  0}$, the first two terms on the right of \eqref{RESID} are
annihilated by $d_z{}^{\!\dagger}$ and $d_w$.  On the other hand, for
the third term in \eqref{RESID} the relation between $\chi_\lambda$
and $\psi_\lambda$ in \eqref{HODGREL} implies 
\begin{equation}
\sum_{\lambda > 0} \*\chi_\lambda(z) \otimes \chi_\lambda(w) \,=\, -e^4 
\sum_{\lambda > 0} \frac{d\psi_\lambda(z) \otimes \*d\psi_\lambda(w)}{\lambda^2}\,.
\end{equation}
The minus sign appears since ${\*^2 = -1}$ on $\Omega^1_\Sigma$.
Thence from \eqref{SPECII} and \eqref{RESID}, the right-hand side of
the corrected commutation relation is given by 
\begin{equation}\label{RESIDII}
\vol_\Sigma \cdot \delta_\Sigma(z,w) - \left(d_z\!\otimes\!
  \*d_w\right)\!G(z,w) = \sum_{j,k=1}^{2g}\left(\SQ^{-1}\right)^{j k}
\Fe_j(z)\otimes\*\Fe_k(w) + \sum_{\lambda > 0}
\chi_\lambda(z)\otimes\*\chi_\lambda(w)\,.
\end{equation}
The expression in \eqref{RESIDII} is manifestly annihilated by
$d_z{}^{\!\dagger}$ and $d_w$, so is compatible with Coulomb gauge.

\medskip\noindent{\sl Maxwell Hilbert Space}\medskip

With the corrected commutation relation in \eqref{CONAII},
the quantization of Maxwell theory on $\Sigma$ is now
straightforward.  From the (slightly formal) functional perspective,
the momentum $\Pi_A$ becomes identified with the operator
\begin{equation}\label{BIGPIA}
\Pi_A(w) \,=\, -i\left[\vol_\Sigma\cdot\frac{\delta}{\delta A(w)} \,+\,
\int_\Sigma d^2u \left(d_u\!\otimes\!
  \*d_w\right)\!G(u,w) \, \vol_\Sigma\cdot\frac{\delta}{\delta A(u)}\right].
\end{equation}
Here the smeared term involving the Green's function $G(u,w)$ is the necessary
price of working in Coulomb gauge.  I do not wish to belabor the interpretation
of the non-local, smeared term, as we will be primarily interested in situations
for which it does not matter.  However, let me say a word about the
basic geometric meaning of \eqref{BIGPIA}, which may be somewhat opaque.

Because ${\Pi_A\in\Omega^1_\Sigma}$ transforms as a one-form, the
right-hand side of \eqref{BIGPIA} must also transform 
as a one-form on $\Sigma$.  Dually to $A$, the derivative
${\delta/\delta A \in T\Sigma}$ transforms as a vector field on
$\Sigma$.  The notation ${\vol_\Sigma \cdot \delta/\delta A}$
indicates that this vector field is to be contracted with the volume
form to produce a one-form on $\Sigma$, as expressed in local
coordinates 
\begin{equation}
\vol_\Sigma \cdot \frac{\delta}{\delta A} \,=\, \sum_{\mu,\nu=1}^2
\left(\vol_\Sigma\right)_{\mu \nu} \, dx^\mu \, \frac{\delta}{\delta
  A_\nu} \,\in\,\Omega^1_\Sigma\,.
\end{equation}
For the smeared term in \eqref{BIGPIA}, we take the wedge-product of
${\left(d_u\!\otimes\! \*d_w\right)\!G(u,w)}$ as a section of
$\Omega^1_\Sigma\otimes\Omega^1_\Sigma$ with the one-form ${\vol_\Sigma \cdot
  \delta/\delta A}$ to obtain a section of ${\Omega^2_\Sigma \otimes
  \Omega^1_\Sigma}$.  The first factor is integrated over $\Sigma$ to
produce yet another one-form.

The classical relation ${\Pi_A = \*E_A/2\pi e^2
  - \upbeta/2\pi}$ implies that the electric field
$E_A$ acts as the covariant operator 
\begin{equation}\label{DOVDAE}
\frac{1}{2\pi e^2}\,\*E_A(w) \,=\, -i\,\frac{D}{D A(w)}\,,
\end{equation}
where
\begin{equation}\label{DOVDA}
\frac{D}{D A(w)} \,=\, \vol_\Sigma\cdot\frac{\delta}{\delta A(w)}
\,+\, \frac{i\,\upbeta}{2\pi} \,+\, 
\int_\Sigma d^2u \left(d_u\!\otimes\!
  \*d_w\right)\!G(u,w) \, \vol_\Sigma\cdot\frac{\delta}{\delta A(u)}\,.
\end{equation}
This expression for $D/D A(w)$ should be compared to the corresponding
expression for $D/D\phi(w)$ appearing in \eqref{DOVDPHI}.  

Smearing or no, since $A$ itself does not appear on the right-hand
side of \eqref{DOVDA}, the functional derivative $D/DA(w)$ describes a
flat connection on the subspace of the affine space $\CA$ where 
${d{}^{\dagger}A = 0}$,
\begin{equation}
\left[\frac{D}{DA(z)},\,\frac{D}{DA(w)}\right] =\, 0\,.
\end{equation}
In precise analogy to the angle $\theta$ appearing in $D/D\phi(w)$,
the harmonic one-form $\upbeta$ in $D/DA(w)$ will
describe the holonomy of a flat connection over the Jacobian
$\SJ_\Sigma$, after we reduce to zero-modes.

As a more down-to-earth alternative to the functional calculus, 
quantization of $A$ can be carried out in terms of the eigenmode
expansion in \eqref{EIGENA}, coupled with the spectral identity in
\eqref{RESIDII}.  To realize the commutator in \eqref{CONAII}, the
pairs $\left(\varphi_0{}^{\!j},\,p_{0,j}\right)$ for ${j=1,\dots,2g}$
and $\left(a_\lambda,a_\lambda{}^{\!\dagger}\right)$ for all ${\lambda
  > 0}$ are promoted to operators which obey the Heisenberg algebra
\begin{equation}\label{HEISA}
\begin{aligned}
&\left[\varphi_0{}^{\!j},\,p_{0,k}\right] \,=\,
i\,\delta_k{}^{\!j}\,,\qquad\qquad j,k\,=\,1,\ldots,2g\,,\\
&\left[a_\lambda,\,a_{\lambda'}{}^{\!\dagger}\right] \,=\,
\frac{\lambda}{e^2} \, \delta_{\lambda \lambda'}\,,
\end{aligned}
\end{equation}
with all other commutators vanishing.  This algebra is akin to that
for the periodic scalar field $\phi$ in \eqref{HEISI}, but rather than
quantizing a single periodic zero-mode, we quantize a
set of $2g$ zero-modes which describe the motion of a
free particle on the Jacobian $\SJ_\Sigma$ of $\Sigma$.

The free-field algebra holds in each topological sector labelled by
${m = \deg(L)}$, so the Maxwell Hilbert space $\SH^\vee_\Sigma$ is the
direct sum
\begin{equation}
\SH_\Sigma^\vee \,=\, \bigoplus_{m\in\BZ} \, \left(\SH_\Sigma^\vee\right){}^{\!m}\,,
\end{equation}
where each summand is itself the tensor product 
\begin{equation}\label{TENSORPA}
\left(\SH_\Sigma^\vee\right){}^{\!m} \,=\,  {\mathsf H}_0^{\vee}
\otimes \bigotimes_{\lambda > 0} {\mathsf H}_\lambda\,.
\end{equation}
Exactly as in \eqref{TENSORP}, ${\mathsf H}_\lambda$ is the oscillator
Fock space acted upon by the pair 
$(a_\lambda,a_\lambda{}^{\!\dagger})$.  We have already
noted that an identical spectrum of non-zero frequencies ${\lambda > 0}$
occurs for both the periodic scalar field and the $U(1)$ gauge
field on $\Sigma$.  Thus the same tensor
product of Fock spaces ${\mathsf H}_\lambda$ appears in both the
scalar Hilbert space $\SH_\Sigma$ and the Maxwell Hilbert space
$\SH_\Sigma^\vee$.  At least for the excited oscillator states in the
two Hilbert spaces, abelian duality is a trivial equivalence.

The remaining factor ${\mathsf H}_0^\vee$ arises from the quantization
of the zero-modes for the gauge field.  Classically,
$(\varphi_0{}^{j},\,p_{0,j})$ for ${j=1,\ldots,2g}$ are coordinates
on the cotangent bundle $T^*\!\!\SJ_\Sigma$ with its canonical symplectic
structure, so the standard quantization yields 
\begin{equation}\label{LTWOJAC}
{\mathsf H}_0^\vee \,\simeq\,
L^2\!\left(\SJ_\Sigma;\BC\right).\qquad\qquad [\,\upbeta = 0\,]
\end{equation}
Because ${\SJ_\Sigma \simeq U(1)^{2 g}}$ is just a torus, the Hilbert
space ${\mathsf H}_0^\vee$ is naturally spanned by the collection of
Fourier wavefunctions 
\begin{equation}\label{WAVEPSIA}
\Psi{}_\omega(\varphi_0) \,=\,
\exp{\!\left(i\,\sum_{j=1}^{2g}\,\varphi_0{}^{j}\int_\Sigma 
    \Fe_j\^\omega\right)}\,,
\qquad\qquad \omega \,\in\, \BL,
\end{equation}
each labelled by an element $\omega$ in the cohomology
lattice ${\BL=H^1(\Sigma;\BZ)}$.  Again, the coincidence of notation
with the winding-number in Section \ref{Scalar} is no accident.
Here though, integrality of $\omega$ is not due to topology per
  se, but rather to the requirement that 
the wavefunction in \eqref{WAVEPSIA} be invariant under shifts
${\varphi_0{}^j \mapsto \varphi_0{}^j + 2\pi}$ for all ${j=1,\ldots,2g}$.  

Physically, $\omega$ determines the conserved momentum carried by the
state $\Psi_\omega(\varphi_0)$, where we apply the identification
\begin{equation}
{\bf W_j} \,=\, p_{0,j} \,=\,
-i\,\frac{\partial}{\partial\varphi_0{}^{j}}\,.\qquad\qquad
[\,\upbeta = 0\,]
\end{equation}
Directly, the Fourier wavefunction in \eqref{WAVEPSIA} is a momentum
eigenstate,
\begin{equation}
{\bf W_j}\cdot \Psi{}_\omega \,=\,
\langle\Fe_j,\,\omega\rangle \cdot 
\Psi{}_\omega\,.
\end{equation} 
As in \eqref{INTERSECT}, ${\langle\,\cdot\,,\,\cdot\,\rangle}$ is
shorthand for the intersection pairing of one-forms on $\Sigma$.

When the topological parameter ${\upbeta \in \CH^1(\Sigma)}$ is
non-zero, the interpretation of the zero-mode momentum $p_{0,j}$ is
modified via \eqref{DOVDAE} and \eqref{DOVDA} to
\begin{equation}\label{COVDPO}
p_{0,j} \,=\, -i\,\frac{D}{D\varphi_0{}^{j}}\,,\qquad\qquad
\frac{D}{D\varphi_0{}^{j}} \,=\, \frac{\partial}{\partial\varphi_0{}^{j}}
\,+\, i\,\frac{\left\langle\Fe_j,\upbeta\right\rangle}{2\pi}\,.
\end{equation}
In precise analogy to the expression for $D/D\phi_0$ in \eqref{DDTHETA},
$D/D\varphi_0$ is the covariant derivative associated to a unitary
flat connection on a complex line-bundle $\CL$ over the Jacobian
$\SJ_\Sigma$, and the harmonic one-form $\upbeta$ determines the
holonomies of this connection around each one-cycle on the Jacobian.
Because $\SJ_\Sigma$ is the quotient ${H^1(\Sigma;\BR)/2\pi\BL}$, each
generating one-cycle ${C_j \in H_1(\SJ_\Sigma;\BZ)}$ can be identified
with a corresponding lattice generator  ${\Fe_j\in\BL}$, for which 
\begin{equation}\label{HOLCJs}
{\Hol}_{C_j}(D/D\varphi_0) \,=\,
\exp{\!\left[-i\left\langle\Fe_j,\upbeta\right\rangle\right]}\,.
\end{equation}

When ${\upbeta\neq 0}$, the Hilbert space ${\mathsf
  H}_0^\vee$ generalizes to the space of square-integrable sections of the
complex line-bundle $\CL$,
\begin{equation}\label{HZERODL}
{\mathsf H}_0^\vee \,\simeq\, L^2\big(\SJ_\Sigma;\CL\big).
\end{equation}
As a result of the covariant identification in \eqref{COVDPO}, ${{\bf
  W_j} \equiv p_{0,j}}$ then acts on the Fourier basis for ${\mathsf H}_0^\vee$ 
with the new eigenvalues 
\begin{equation}\label{EVALBIGP}
{\bf W_j} \cdot \Psi_\omega \,=\, \left\langle\Fe_j,\,\omega +
  \frac{\upbeta}{2\pi}\right\rangle\cdot\Psi_\omega\,,\qquad\qquad
\omega\,\in\,\BL\,.
\end{equation}
Again in comparison to \eqref{BFPOPII}, the role of the harmonic one-form
$\upbeta$ is to shift the integral grading by the
cohomology lattice $\BL$ on the Maxwell Hilbert space $\SH_\Sigma^\vee$.

Because the zero-mode Hilbert space ${\mathsf H}_0^\vee$ is graded by
the eigenvalues of ${\bf W_j}$ for ${j=1,\ldots,2g}$, the full
Maxwell Hilbert space $\SH_\Sigma^\vee$ is bigraded by the lattice
${\BL\oplus\BZ}$,
\begin{equation}
\SH_\Sigma^\vee \,\simeq \bigoplus_{(\omega,\,m)\in\BL\oplus\BZ}
\left(\SH_\Sigma^\vee\right){}^{\!\omega,m}\,.
\end{equation}
Following the notation in Section \ref{Scalar}, I let
$|\omega;m\rangle$ denote the Fourier wavefunction
$\Psi_\omega(\varphi_0)$, considered in the topological sector with
magnetic flux ${m = \deg(L)}$, and satisfying the vacuum condition
\begin{equation}
a_\lambda|\omega;m\rangle = 0\,,\qquad\qquad \lambda
  > 0\,.
\end{equation}
All other Fock states in $\SH_\Sigma$ are obtained by acting with the
oscillator raising-operators $a_\lambda{}^{\!\dagger}$ on the Fock vacuum
$|\omega;m\rangle$, so more explicitly 
\begin{equation}\label{CURLYHSV}
\left(\SH_\Sigma^\vee\right){}^{\!\omega,m} =\,
\BC\cdot|\omega;m\rangle \otimes \bigotimes_{\lambda > 0}
{\mathsf H}_\lambda\,.
\end{equation}
Clearly $\left(\SH_\Sigma^\vee\right){}^{\!\omega,m}$ is isomorphic to
the scalar field summand $\SH_\Sigma^{\,m,\omega}$ in \eqref{CURLYHS}.

Finally, let us consider the action of the Maxwell Hamiltonian ${\bf
  H}^\vee$ on states in the Hilbert space $\SH_\Sigma^\vee$.  Under the
identification \eqref{DOVDAE} of the electric field $\*E_A$ with the
covariant operator $D/DA$, the Hamiltonian becomes  
\begin{equation}
{\bf H}^\vee \,=\, \int_\Sigma \left[-\pi e^2\frac{D}{D A}\^\*\frac{D}{D
    A}\, +\, \frac{1}{4\pi e^2} \,F_A\^\*F_A 
  \,-\, \frac{\theta}{2\pi e^2 \ell^2}\,F_A\right].
\end{equation}
In terms of the conserved momenta ${\bf W_j}$ in
\eqref{BIGDULP},
\begin{equation}
{\bf H}^\vee \,=\,\int_\Sigma \left[\pi e^2\left(\SQ^{-1}\right){}^{\!j k} \, {\bf
    W_j}\,{\bf W_k}\,+\, \cdots \,+\, \frac{1}{4\pi e^2} \,F_A\^\*F_A 
  \,-\, \frac{\theta}{2\pi e^2 \ell^2}\,F_A\right],
\end{equation}
where the omitted terms involve the action of $D/DA$ on the excited
oscillator states in the Hilbert space.

The complete spectrum of the Maxwell Hamiltonian depends upon the set
of eigenvalues $\{\lambda^2\}$ for the scalar Laplacian on $\Sigma$,
exactly as for the periodic scalar field.  Following the strategy in
Section \ref{Scalar}, we ask instead the 
more limited question of how ${\bf H}^\vee$ acts on the Fock  
vacua $|\omega;m\rangle$ associated to the harmonic modes of the gauge
field.  Evidently from \eqref{FHAT} and \eqref{EVALBIGP},
\begin{equation}\label{SPECHM}
{\bf H}^\vee|\omega;m\rangle \,=\, e^2\left[\pi\!\left(\omega +
    \frac{\upbeta}{2\pi},\,\omega + \frac{\upbeta}{2\pi}\right) +\,
  \frac{\pi\,m^2}{\left(e^2\ell\right)^2} \,-\,
  \frac{\theta\,m}{\left(e^2\ell\right)^2} \,+\,
  \frac{E_0}{e^2\ell}\right]\! |\omega;m\rangle\,,
\end{equation}
where $E_0/\ell$ is again a Casimir energy on $\Sigma$.  Because the
zero-point energies $\ha\lambda$ of the oscillating modes are the same
for both the periodic scalar field and the gauge field, the constant
$E_0$ in \eqref{SPECHM} will agree with the corresponding constant in
\eqref{SPECTH} so long as we use the same regularization method to 
define both (as we assume).

For Maxwell theory on $\Sigma$, the spectrum of ${\bf H}^\vee$
simplifies in the regime ${e^2\ell \ll 1}$ of  weak electric coupling.
Only then do the Fock vacua $|\omega;m\rangle$ for arbitrary Fourier momentum
${\omega\in\BL}$ have parametrically smaller energy than the typical
oscillator state such as $a_\lambda{}^{\!\dagger}|0;m\rangle$.  Not
surprisingly, this case is opposite to the strong-coupling regime
${1/e^2\ell \ll 1}$ in which the states of least-energy arise by
quantizing the single zero-mode of the periodic scalar field $\phi$.

\subsection{Topological Hilbert Space}\label{Duality}

Let us summarize our results so far.  

We have obtained an explicit identification between the Hilbert spaces
for the $U(1)$ gauge field $A$ and the periodic scalar field $\phi$ on
the surface $\Sigma$,
\begin{equation}\label{DUALIDH}
\begin{aligned}
\SH_\Sigma{}^{\!\vee} \,&\simeq\, \SH_\Sigma\,,\\
&=\, \bigoplus_{(m,\,\omega)\in\BZ\oplus\BL}
\left[\BC\cdot|m;\omega\rangle \otimes \bigotimes_{\lambda > 0}
{\mathsf H}_\lambda\right].  
\end{aligned}
\end{equation}

The isomorphism for the oscillator Fock spaces ${\mathsf H}_\lambda$
for ${\lambda > 0}$ follows from classical Hodge theory after we pass
to Coulomb gauge for $A$, so it is relatively uninteresting.
The non-trivial content in \eqref{DUALIDH} is the identification between
the Fock vacua $|m;\omega\rangle$, which arise from the quantization of the
harmonic modes of $A$ and $\phi$ in each topological sector.  

For the periodic scalar field, ${m\in\BZ}$ is a quantum label which
arises from Fourier modes on $S^1$, and the lattice vector 
${\omega\in\BL}$ is a classical label which measures the
winding-number of the map ${\phi:\Sigma\to S^1}$.  Conversely for the
gauge field, the integer $m$ is the classical label, corresponding to the degree of the
line-bundle $L$, and the lattice vector $\omega$ is the quantum 
label, arising from Fourier modes on the Jacobian $\SJ_\Sigma$.  
Under the isomorphism in \eqref{DUALIDH}, the classical and quantum
labels are swapped, characteristic of abelian duality in any dimension.

A dual role is also played by the topological parameters
$(\theta,\upalpha)$ and $(\theta,\upbeta)$ which enter the respective
Hamiltonians in \eqref{CLASSH} and \eqref{BIGHA}.  For the
periodic scalar field, the angle $\theta$ is a quantum parameter which
determines the holonomy of a flat, unitary connection on a complex line
bundle over $S^1$ as in \eqref{HOLSONE}, and the harmonic one-form
$\upalpha$ is a classical parameter which weights each winding-sector.
For the gauge field, $\theta$ is the classical parameter which weights the 
magnetic flux on $\Sigma$, and $\upbeta$ is now the quantum
parameter which determines the holonomy of a flat connection on
a complex line-bundle over $\SJ_\Sigma$ as in \eqref{HOLCJs}.

Nonetheless, under the dual correspondence
\begin{equation}
\upalpha \,=\, \*\upbeta\,,\qquad\qquad
\upalpha,\upbeta\,\in\,\CH^1(\Sigma)\,,
\end{equation}
the Hamiltonians ${\bf H}$ in \eqref{SPECTH} and ${\bf H}{}^{\vee}$ in
\eqref{SPECHM} act identically on the states $|m;\omega\rangle$ up to 
a constant shift $\delta$,
\begin{equation}\label{SHIFTH}
{\bf H}|m;\omega\rangle \,=\, \Big({\bf H}^\vee \,+\,
  \delta\Big)|m;\omega\rangle\,,\qquad \delta =
  \frac{e^2}{4\pi}\left[\frac{\theta^2}{ 
    (e^2\ell)^2} \,-\, \left(\upbeta,\upbeta\right)\right]\,.
\end{equation}
Let us introduce the quantum partition functions for the scalar and
Maxwell theories,
\begin{equation}
Z_\Sigma(R) \,=\, \Tr_{\SH_\Sigma}\,\e{-R\,{\bf H}}\,,\qquad\qquad
Z_\Sigma{}^{\!\!\vee}(R) \,=\, \Tr_{\SH_\Sigma}\,\e{-R\,{\bf H}^\vee}\,,
\end{equation}
both of which depend upon a real parameter ${R\in\BR}$ which can be
interpreted as the length of the circle in ${M = S^1 \times \Sigma}$.

The constant shift in \eqref{SHIFTH} then implies the relation 
\begin{equation}\label{DUALRZ}
Z_\Sigma{}^{\!\!\vee}(R) \,=\, Z_\Sigma(R) \cdot
\exp{\!\left[\frac{e^2 R}{4\pi}\left(\frac{\theta^2}{  
    (e^2\ell)^2} \,-\, \left(\upalpha,\upalpha\right)\right)\right]}\,.
\end{equation}
The same duality relation appears under a different guise in
\cite{BeasleyI:2014}, where it arises from the non-trivial
modular transformation of a theta-function $\Theta_M(\upgamma)$
associated to any Riemannian three-manifold $M$. See Section $4.1$ of
\cite{BeasleyI:2014} for a complete discussion of the theta-function
and Section $5.1$ of the same work for a path integral derivation of
the relation in \eqref{DUALRZ}.  Compare especially to equation
$(5.1)$ in \cite{BeasleyI:2014}.

Finally, as we have already mentioned, the spectrum of ${\bf H}$
dramatically simplifies in either the small-volume limit ${e^2\ell \ll
  1}$ or the large-volume limit ${e^2 \ell \gg 1}$.  In both cases,
the quantum states of minimal energy within each topological sector
are the Fock vacua $|m;\omega\rangle$, for all pairs $(m,\omega)$ in
the lattice ${\BZ\oplus\BL}$.  Hence we can sensibly restrict
attention to the subspace of the full Hilbert space 
spanned by these states,
\begin{equation}
\SH_\Sigma^{\rm top} = \bigoplus_{(m,\omega)\in\BZ\oplus\BL}
\BC\cdot|m;\omega\rangle \, \subset\, \SH_\Sigma\,.
\end{equation}

Essential for the following, the description of $\SH_\Sigma^{\rm top}$
does not require detailed knowledge of the Riemannian metric on
$\Sigma$.  Instead, the action of operators such as ${\bf H}$ on
$\SH_\Sigma^{\rm top}$ will only depend upon the  
complex structure and the overall volume of $\Sigma$.
In that sense, $\SH_\Sigma^{\rm top}$ is a subspace of
`quasi-topological' states.   Unlike the typical situation in
topological quantum field theory, though, $\SH_\Sigma^{\rm top}$ has
infinite dimension.  As a result, the action of various operators on
$\SH_\Sigma^{\rm top}$ can be quite interesting, a topic to which we
turn next.  Elsewhere, I will discuss some important related notions in the context
of ${\CN=2}$ supersymmetric quantum field theory in three dimensions.

\medskip
\section{Operator Algebra at Higher Genus}\label{Algebra}

Given the explicit construction of the Hilbert space on $\Sigma$, we
now discuss the action of several natural classes of operators on that Hilbert
space.

As mentioned at the end of Section
\ref{Duality}, we simplify life by considering only the action on
the quasi-topological subspace ${\SH_\Sigma^{\rm top}}$ spanned by the
Fock vacua $|m;\omega\rangle$,
\begin{equation}
\SH_\Sigma^{\rm top} = \bigoplus_{(m,\,\omega)\in\BZ\oplus\BL}
\BC\cdot|m;\omega\rangle\,.
\end{equation}
The restriction to $\SH_\Sigma^{\rm top}$ is natural in either the
regime ${e^2\ell \ll 1}$ or ${e^2 \ell \gg 1}$, for which the Fock
vacua describe states of minimal energy within each topological
sector.  Because the abelian theories in question are non-interacting, we do
not need to worry about the effects of high-energy states, which would
otherwise be integrated-out in passing from the big Hilbert space
$\SH_\Sigma$ to the subspace $\SH_\Sigma{}^{\!\rm top}$.

Following Section $5.2$ in \cite{BeasleyI:2014}, we analyze three classes
of operators on $\Sigma$.  We first have the local vertex operator
$\SV_k(\sigma)$ which is inserted at a point ${\sigma \in \Sigma}$,
\begin{equation}\label{VERTEXP}
\SV_k(\sigma) \,=\, \e{\! i k\phi(\sigma)}\,,\qquad\qquad  k \,\in\,\BZ\,.
\end{equation}
Periodicity of the scalar field ${\phi \sim \phi + 2\pi}$ dictates
that $k$ be an integer so that $\SV_k(\sigma)$ is
single-valued.

Next we have the Wilson loop operator $\SW_n(C)$ associated to an
oriented, smoothly embedded curve ${C \subset \Sigma}$,
\begin{equation}\label{WILSONII}
\SW_n(C) \,=\, \exp{\!\left[i\,n\oint_C A\right]}\,,\qquad\qquad 
n\,\in\,\BZ\,.
\end{equation}
For generic choices of $C$, the charge $n$ of the Wilson loop operator
must be an integer to ensure gauge-invariance with respect to the compact
gauge group $U(1)$.  

Perhaps less appreciated, when $C$ is a homologically-trivial
curve which bounds a two-cycle ${D \subset \Sigma}$, the Wilson loop operator
can be defined for an arbitrary real charge via 
\begin{equation}\label{WILSONNU}
\SW_\nu(C) \,=\, \exp{\!\left[i\,\nu\int_D
    F_A\right]}\,,\qquad\qquad \nu\,\in\,\BR\,.
\end{equation}
This expression for $\SW_\nu(C)$ is manifestly gauge-invariant for all
values of $\nu$, and it reduces to \eqref{WILSONII} by Stokes' theorem
for ${C = \partial D}$.  Unlike the situation for $\SW_\nu(C)$ in
three dimensions \cite{BeasleyI:2014}, where the role of $D$ is played
by a Seifert surface with some homological ambiguity, here there is
no ambiguity about $D$. Because $D$ is a two-cycle on $\Sigma$, the choice
of $D$ is fixed entirely by the orientations of the pair $(\Sigma,
C)$.  

Without delay, let me emphasize that $\SW_\nu(C)$ will act
non-trivially on $\SH_\Sigma^{\rm top}$ even when $C$ is trivial in
homology.  Likewise, $\SW_n(C)$ will depend upon the geometry of ${C \subset
  \Sigma}$, not just the homology class ${[C]\in H_1(\Sigma)}$.

By contrast, we do have a purely homological loop operator
$\SL_{\alpha}(C)$, given by 
\begin{equation}\label{LOPPPHI}
\SL_{\alpha}(C) \,=\, \exp{\!\left[\frac{i\,\alpha}{2\pi}\oint_C
    d\phi\right]}\,,\qquad\qquad \alpha\,\in\,\BR/2\pi\BZ\,.
\end{equation}
Clearly $\SL_\alpha(C)$ detects the classical winding-number of
the map ${\phi:\Sigma\to S^1}$, for which only the homology class ${[C]\in
  H_1(\Sigma)}$ is relevant.  Since the one-form $d\phi/2\pi$ always has
integral periods, $\SL_\alpha(C)$ also depends only upon the value of
$\alpha$ modulo $2\pi$.

The operators in \eqref{VERTEXP}, \eqref{WILSONII}, and
\eqref{LOPPPHI} are presented in order-form, as
classical functionals of the scalar field $\phi$ or the gauge field
$A$. As well-known and reviewed for instance in Section $5.2$ of
\cite{BeasleyI:2014}, each of these operators admits a dual disorder
description, in which the operator creates a classical singularity at
the point $\sigma$ or along the curve $C$, respectively.

Very briefly, the vertex operator $\SV_k(\sigma)$ creates a local monopole singularity of
magnetic charge $k$ in the gauge field $A$, and the loop operator
$\SL_{\alpha}(C)$ creates a codimension-two singularity in ${M =
  \BR \times \Sigma}$ around which $A$ has monodromy $\alpha$.  Note
that this interpretation is consistent with the angular nature of
$\alpha$.  For the periodic scalar field, the Wilson loop operator 
$\SW_n(C)$ dually creates an additive monodromy ${\phi \mapsto \phi +
  2\pi n}$ around any small path encircling $C$ inside $M$.  Note
that integrality of $n$ is required for the monodromy to make sense
for general $C$.

Such classical geometric descriptions of the disorder operators
suffice for the path integral analysis of duality in
\cite{BeasleyI:2014}.  Our goal in Section \ref{Monopoles} is to
provide an alternative, quantum description of these operators -- in both
order and disorder form -- by their action on the Hilbert space
$\SH_\Sigma{}^{\!\rm top}$.  Using these results, we then exhibit
directly in Section \ref{Heisenberg} the combined algebra of vertex and loop
operators on $\Sigma$.

\subsection{Monopoles and Loops on a Riemann Surface}\label{Monopoles}

To discuss the action of $\SV_k(\sigma)$, $\SW_n(C)$, and
$\SL_\alpha(C)$ on the Hilbert space, we assume that each
operator acts at time ${t = 0}$ on ${M = \BR  \times \Sigma}$.
These operators do not generally preserve the topological subspace
$\SH_\Sigma^{\rm top}$ inside the full Hilbert space $\SH_\Sigma$.   To obtain an action on
$\SH_\Sigma^{\rm top}$ alone, we compose with the
projection from $\SH_\Sigma^{}$ onto $\SH_\Sigma^{\rm top}$, which
occurs naturally in either the geometric limits ${e^2\ell \ll 1}$ or ${e^2
  \ell \gg 1}$.  This projection onto $\SH_\Sigma^{\rm top}$ will be
implicit throughout.  At the classical level, projection onto
$\SH_\Sigma^{\rm top}$ amounts to the Hodge projection onto harmonic
configurations for $\phi$ and $A$.

\medskip\noindent{\sl Monopole Operators}\medskip

We begin with the action of the vertex operator $\SV_k(\sigma)$ in
\eqref{VERTEXP}.  Directly via the eigenmode expansion \eqref{EIGENOM} for the periodic scalar
field, 
\begin{equation}\label{VSONE}
\SV_k(\sigma)|m;\omega\rangle \,=\, 
\exp{\!\left[i\,k\,\Phi_\omega(\sigma) \,+\, i\,\frac{k}{e^2 \ell}\,\phi_0
    \,+\, \cdots \right]}|m;\omega\rangle\,,\qquad
(m,\omega) \in \BZ\oplus\BL\,.
\end{equation}
Here ${\Phi_\omega:\Sigma\to S^1}$ is the fiducial harmonic map
\eqref{HARM1} with winding-number $\omega$, and the ellipses indicate
terms involving the Fock operators $a_\lambda$ and
$a_\lambda{}^{\dagger}$, whose action becomes irrelevant after the projection
to $\SH_\Sigma^{\rm top}$.  

According to the description of the
Fourier wavefunction in \eqref{FOURIER}, the Fock groundstate $|m;\omega\rangle$
can itself be written as 
\begin{equation}
|m;\omega\rangle \,\equiv\, \Psi_m(\phi_0)|\omega\rangle \,=\,
\exp{\!\left(i\,\frac{m}{e^2\ell}\,\phi_0\right)}|\omega\rangle\,.
\end{equation}
Evidently from \eqref{VSONE}, $\SV_k(\sigma)$ shifts the Fourier mode
number $m$ to ${m+k}$,
\begin{equation}\label{VSTWO}
\SV_k(\sigma)|m;\omega\rangle \,=\,
\exp{\!\Big[i\,k\,\Phi_\omega(\sigma)\Big]} |m+k;\omega\rangle\,,
\end{equation}
up to an additional phase which depends upon the value of $\Phi_\omega$
at the point $\sigma$.  When ${\Sigma = \BC\BP^1}$ has genus zero, the harmonic map
${\Phi_{\omega}}$ in \eqref{VSTWO} is constant and equal to zero
modulo $2\pi$ by the defining condition in \eqref{HARM2}.  Hence in this case, the action of
$\SV_k(\sigma)$ on the quasi-topological subspace $\SH_\Sigma^{\rm
  top}$ does not actually depend upon the position of the vertex
operator on $\Sigma$.  

In higher genus, the situation is more interesting.  

To evaluate the
phase in \eqref{VSTWO}, we use the defining conditions for the
fiducial map, namely 
\begin{equation}
d\Phi_\omega \,=\, 2 \pi \omega\,,\qquad\qquad \Phi_\omega(\sigma_0)
\,=\, 0 \,\,\mod\,\, 2\pi\,,
\end{equation}
where ${\omega\in H^1(\Sigma;\BZ)}$ is harmonic and
${\sigma_0\in\Sigma}$ is the basepoint used for quantization.  By
Stokes' theorem, the phase factor in \eqref{VSTWO} can be recast in the form 
\begin{equation}\label{VSTHRE}
\begin{aligned}
\exp{\!\Big[i\,k\,\Phi_\omega(\sigma)\Big]} \,&=\, \exp{\!\Big[i\, k
    \big(\Phi_\omega(\sigma) \,-\,
      \Phi_\omega(\sigma_0)\big)\Big]}\,,\\
&=\, \exp{\!\Big[2\pi i\, k \int_\Gamma
  \omega\Big]}\,,
\end{aligned}
\end{equation}
where $\Gamma$ is any oriented path on $\Sigma$ which connects the basepoint
$\sigma_0$ to the point $\sigma$ where the vertex operator is inserted,
\begin{equation}\label{BDYGAM}
\partial\Gamma \,=\, \sigma \,-\, \sigma_0\,.
\end{equation}
The homotopy class of $\Gamma$ is not unique, as clear when ${\sigma =
  \sigma_0}$ and $\Gamma$ is an arbitrary closed curve based at
$\sigma_0$.  However, as usual in the business, integrality of both
$k$ and $\omega$ ensures that the phase in \eqref{VSTHRE} is
independent of the choice of the integration contour 
$\Gamma$.  For the remainder, we suppress the appearance of $\Gamma$ and
simply write the vertex operator phase as 
\begin{equation}\label{VSTOP}
\SV_k(\sigma)|m;\omega\rangle \,=\,
\exp{\!\Big[2\pi i\,k \int_{\sigma_0}^\sigma\omega\Big]} |m+k;\omega\rangle\,.
\end{equation}
Thus, even when we restrict to the low-energy subspace
${\SH_\Sigma^{\rm top} \subset \SH_\Sigma^{}}$, the action of the vertex operator
$\SV_k(\sigma)$ is still sensitive to the location at which the
operator is inserted.

How does \eqref{VSTOP} arise when we describe
the quantum theory on $\Sigma$ dually in terms of the Maxwell gauge
field $A$?  To answer this question, we recall that the Fock vacua
$|m;\omega\rangle$ in $\SH_\Sigma^{\rm top}$ correspond 
to wavefunctions for $A$ on the disjoint union of tori 
\begin{equation}
\Pic(\Sigma) =
  \bigsqcup_{m\in\BZ} \Pic_m(\Sigma)\,,\qquad\qquad \Pic_m(\Sigma)
  \simeq \SJ_\Sigma\,,
\end{equation}
each isomorphic to the Jacobian of $\Sigma$.  A natural guess is that
the effective action of the vertex operator $\SV_k(\sigma)$ is induced
from the tensor product (or Hecke modification) with the degree-$k$
holomorphic line-bundle $\CO_\Sigma(k\,\sigma)$,
\begin{equation}\label{HECKE}
\begin{aligned}
\otimes\CO_\Sigma(k\,\sigma):\Pic_m(\Sigma)
\,&\buildrel\simeq\over\longrightarrow\,
\Pic_{m+k}(\Sigma)\,,\\
{\mathfrak L} \quad&\longmapsto\quad\!\! {\mathfrak L}\otimes
\CO_\Sigma(k\,\sigma)\,,
\end{aligned}
\end{equation}
as already appears in \eqref{ISOPIC}.  In the gauge theory approach to
geometric Langlands, this statement has been explained in
\S $9.1$ of \cite{Kapustin:2006pk}, though we must make a few minor
modifications to treat the non-topological (but free) theory here.

According to its definition as a monopole operator, reviewed in
Section $5.2$ of \cite{BeasleyI:2014}, the vertex operator
$\SV_k(\sigma)$ acts topologically to increase the degree of the
$U(1)$-bundle over $\Sigma$ by $k$ units, in accord with the
shift ${m \mapsto m+k}$ in both \eqref{VSTOP} and \eqref{HECKE}.
However, for a complete characterization of the phase in \eqref{VSTOP}, we must also consider how
the action of $\SV_k(\sigma)$ on the state $|m;\omega\rangle$ depends upon
the point ${\sigma\in\Sigma}$ and the Fourier mode ${\omega\in\BL}$
for the gauge field wavefunction on $\SJ_\Sigma$.  

To investigate the latter dependence, let us consider the composite
operator 
\begin{equation}
\SO_k(\sigma,\sigma_0) \,=\, \SV_k(\sigma) \circ
\SV_{-k}(\sigma_0)\,,\qquad\qquad \sigma \neq \sigma_0\,,
\end{equation}
where $\sigma$ is distinct from the basepoint $\sigma_0$.  Because
$\SV_k(\sigma)$ and $\SV_{-k}(\sigma_0)$ carry 
opposite monopole charges, ${\SO_k(\sigma,\sigma_0)}$ does not alter the topology of the
line-bundle over $\Sigma$.  Nonetheless, $\SO_k(\sigma,\sigma_0)$
may still act non-trivially on the state $|m;\omega\rangle$, at least in genus
${g \ge 1}$.

Since we work with zero-modes, let $A$ be an arbitrary harmonic connection on a
line-bundle of degree $m$ over $\Sigma$, of the form
\begin{equation}\label{SOMEA}
A \,=\, m\,\widehat{A} \,+\, \sum_{j=1}^{2 g}
\varphi_0{}^j\,\Fe_j\,,\qquad\qquad \varphi_0{}^j \,\in\,\BR/2\pi\BZ\,.
\end{equation}
Here we have truncated the more general eigenform expansion for $A$ in
\eqref{EIGENA}, and we recall that $\widehat{A}$ is the fiducial harmonic
connection associated to the holomorphic line-bundle $\CO_\Sigma(\sigma_0)$.
As a harmonic connection, $A$ determines a point in the Picard
component $\Pic_m(\Sigma)$ of degree $m$, and we may 
interpret the Fock  groundstate 
\begin{equation}
|m;\omega\rangle \,\equiv\, \Psi_\omega(A) |m\rangle
\end{equation}
in terms of the wavefunction 
\begin{equation}\label{EAP}
\Psi_\omega(A) \,=\, \exp{\!\left[i\int_\Sigma \left(A -
      m\,\widehat{A}\,\right)\!\^\omega\right]}\,,
\end{equation}
exactly as in \eqref{WAVEPSIA}.  After we subtract ${m\,\widehat{A}}$
in the argument of the exponential, $\Psi_\omega(A)$ does not actually
depend upon the degree $m$.

On this wavefunction, the composite operator
${\SO_k(\sigma,\sigma_0)}$ acts via a modification of $A$ induced from 
\eqref{HECKE},
\begin{equation}\label{COMPVI}
\SO_k(\sigma,\sigma_0)\cdot\Psi_\omega(A) \,=\,
\Psi_\omega(\wt A)\,,
\end{equation}
where 
\begin{equation}\label{HECKA}
\wt A \,=\, A \,+\, 2 \pi k \, \delta_\Gamma\,,\qquad\qquad\qquad
\delta_\Gamma \,\in\, \Omega^1_\Sigma\,.
\end{equation}
In this expression, $\delta_\Gamma$ is a one-form on $\Sigma$ with delta-function
support which represents the Poincar\'e dual of an oriented path
$\Gamma$ running from $\sigma_0$ to $\sigma$.   Equivalently, for any
smooth one-form $\eta$, the wedge product with $\delta_\Gamma$ satisfies 
\begin{equation}\label{PDGAM}
\int_\Sigma \delta_\Gamma \^ \eta \,=\, \int_\Gamma
\eta\,,\qquad\qquad \eta\,\in\,\Omega^1_\Sigma\,.
\end{equation}
Because the path $\Gamma$ is open, the one-form
$\delta_\Gamma$ is not closed but rather obeys 
\begin{equation}\label{MONPD}
d\delta_\Gamma \,=\, \delta_\sigma -
\delta_{\sigma_0}\,,\qquad\qquad \delta_\sigma, \delta_{\sigma_0}
\,\in\,\Omega^2_\Sigma\,,
\end{equation}
where $\delta_\sigma$ and $\delta_{\sigma_0}$ are two-forms on
$\Sigma$ with delta-function support at the points ${\sigma, \sigma_0 \in \Sigma}$.  As a
consequence of \eqref{MONPD}, the curvature of the modified connection $\wt
A$ in \eqref{HECKA} is singular at the locations where the vertex
operators are inserted.  Physically, these curvature singularities signal the
creation of a monopole/anti-monopole pair of magnetic charge $k$ on
$\Sigma$.

Strictly speaking, the singular connection $\wt A$ in \eqref{HECKA} is not harmonic,
and so to interpret the dual action of ${\SO_k(\sigma,\sigma_0)}$ on
the zero-mode wavefunction, we should project the singular connection $\wt A$ onto the harmonic
subspace of $\Omega^1_\Sigma$.  Thankfully, this projection is
accomplished automatically for us when we evaluate
\begin{equation}\label{COMPVII}
\begin{aligned}
\Psi_\omega(\wt A) \,&=\, \exp{\!\left[ i \int_\Sigma \left(A \,+\,
    2\pi k \,\delta_\Gamma \,-\, m \, \widehat A\,\right)\!\^\omega\right]},\\
&=\, \exp{\!\left( 2 \pi i k  \int_\Sigma
    \delta_\Gamma\^\omega\right)} \cdot \Psi_\omega(A)\,,\\
&=\, \exp{\!\left( 2 \pi i \, k \int_\Gamma \omega\right)} \cdot
\Psi_\omega(A)\,,
\end{aligned}
\end{equation}
since the one-form $\omega$ is harmonic by assumption.  In passing
from the second to the third line of \eqref{COMPVII}, we use the
defining property of $\delta_\Gamma$ in \eqref{PDGAM}.

According to \eqref{COMPVI} and \eqref{COMPVII}, 
\begin{equation}\label{COMPVIV}
\SO_k(\sigma,\sigma_0)|m;\omega\rangle \,=\, 
\Big[\SV_k(\sigma) \circ \SV_{-k}(\sigma_0)\Big]|m;\omega\rangle \,=\,
\exp{\!\left( 2 \pi i \, k \int^\sigma_{\sigma_0} \omega\right)}
|m;\omega\rangle\,,
\end{equation}
and again integrality of both $k$ and $\omega$ ensures that the phase
factor depends only upon the endpoints of the path $\Gamma$, where the
monopoles are inserted.  Clearly from its definition \eqref{HECKE} via
the tensor product, the monopole operator of charge $k$ is the same as the
$k$-th power of the unit monopole operator,
\begin{equation}
\SV_k(\sigma) \,=\, \SV_1(\sigma)^k\,,
\end{equation}
so we can rewrite the identity in \eqref{COMPVIV} as 
\begin{equation}\label{COMPOIV}
\SV_k(\sigma) |m;\omega\rangle \,=\, \exp{\!\left( 2 \pi i \, k
    \int^\sigma_{\sigma_0}\omega\right)} \SV_k(\sigma_0) |m;\omega\rangle.
\end{equation}
Hence we have determined the action of the monopole operator
$\SV_k(\sigma)$ for arbitrary points ${\sigma\in\Sigma}$ in terms of the action
of the monopole operator $\SV_k(\sigma_0)$ inserted at the basepoint
$\sigma_0$.

We are left to discuss the action of the based monopole
$\SV_k(\sigma_0)$ on $|m;\omega\rangle$.  The first 
claim is that $\SV_k(\sigma_0)$ does not alter the quantum label
$\omega$, 
\begin{equation}\label{SOMECH}
\langle m+k;\omega'|\SV_k(\sigma_0)|m;\omega\rangle \,=\,
0\,,\qquad\qquad \omega \,\neq\,\omega'\,.
\end{equation}
This statement follows by symmetry, since $\omega$ is interpreted as
the charge under the group $U(1)^{2 g}$ which acts by
translations on the Jacobian $\SJ_\Sigma$.  For the harmonic
connection $A$ in \eqref{SOMEA}, these translations are just shifts in
the angular coordinates $\varphi_0{}^{j}$.  Because the Hecke
modification in \eqref{HECKE} commutes with the action of $U(1)^{2g}$,
the monopole operator is uncharged under $U(1)^{2g}$ and hence preserves $\omega$.

Otherwise, from the gauge theory perspective we have left some
ambiguity in the normalization of the basis state 
$|m;\omega\rangle$ for fixed $\omega$ as the degree $m$ ranges over $\BZ$.
We fix this ambiguity up to an overall constant by declaring 
\begin{equation}\label{BAS} 
|m;\omega\rangle \,\equiv\, \big[\SV_1(\sigma_0)\big]^m |0;\omega\rangle\,,
\end{equation}
so that 
\begin{equation}\label{BASII}
\SV_k(\sigma_0) |m;\omega\rangle \,=\, |m+k;\omega\rangle\,.
\end{equation}
Together, \eqref{COMPOIV} and \eqref{BASII} imply the formula in
\eqref{VSTOP}, which we deduced from the more direct description of
$\SV_k(\sigma)$ as a vertex operator for the periodic scalar field $\phi$.

\medskip\noindent{\sl Vortex Loops}\medskip

Just as we consider the action by the local vertex operator
$\SV_k(\sigma)$, we can also consider the action on $\SH_\Sigma^{\rm
  top}$ by the respective loop operators $\SL_\alpha(C)$ and
$\SW_n(C)$, where $C$ is a closed curve in $\Sigma$.  If $C$ is not a spacelike curve in
$\Sigma$ but a timelike curve in $M$ of the form ${C = \BR \times
  \{\sigma\}}$ for some point ${\sigma \in \Sigma}$, then
$\SL_\alpha(C)$ and $\SW_n(C)$ do not act on the Hilbert space
$\SH_\Sigma^{\rm top}$ but lead rather to the construction of new
Hilbert spaces associated to the punctured surface ${\Sigma^o = \Sigma
  - \{\sigma\}}$.  The analysis of such line operators is 
similar philosophically to the analysis of fibrewise Wilson loop operators in
\cite{Beasley:2009mb}, so I omit the timelike case here.

Like the vertex operator $\SV_k(\sigma)$, the loop operator
$\SL_\alpha(C)$ admits an elementary description in terms of the
periodic scalar field $\phi$,
\begin{equation}\label{LINEII}
\SL_\alpha(C) \,=\, \exp{\!\left(\frac{i \, \alpha}{2\pi} \oint_C
d\phi\right)}\,,\qquad\qquad \alpha \,\in\, \BR/2\pi\BZ\,.
\end{equation}
On each state $|m;\omega\rangle$ in $\SH_\Sigma^{\rm top}$, the operator $\SL_\alpha(C)$ simply
measures the winding-number ${\omega \in H^1(\Sigma;\BZ)}$,
\begin{equation}\label{LINEAC}
\begin{aligned}
\SL_\alpha(C) |m;\omega\rangle \,&=\, \exp{\!\left(\frac{i \,
      \alpha}{2\pi} \oint_C d\phi\right)} |m;\omega\rangle\,,\\
&=\, \exp{\!\left(i \, \alpha \oint_C \omega\right)}
|m;\omega\rangle\,,
\end{aligned}
\end{equation}
where we recall that ${[d\phi] = 2 \pi \omega}$ for the state
$|m;\omega\rangle$.  In particular, $\SL_\alpha(C)$ respects the
global $U(1)$ symmetry by shifts ${\phi \mapsto \phi + c}$ for
constant $c$ and hence preserves the mode number $m$ of the state
$|m;\omega\rangle$.  We also note that the phase in \eqref{LINEAC}
depends only on the homology class of $C$ in $H_1(\Sigma)$, and the
integrality of $\omega$ ensures that the phase depends only on the
value of the parameter $\alpha$ modulo $2\pi$.  

Again, our main goal is to understand how the formula in
\eqref{LINEAC} arises dually in terms of the $U(1)$ gauge field $A$.
As reviewed in Section $5.2$ of \cite{BeasleyI:2014}, the loop
operator $\SL_\alpha(C)$ acts on $A$ as a disorder operator which
creates a curvature singularity along $C$ of the form  
\begin{equation}\label{SURFAII}
F_A \,=\, -\alpha \, \delta_C\,,\qquad\qquad \alpha \,\in\,\BR/2\pi\BZ\,,
\end{equation}
where $\delta_C$ is a two-form with delta-function support that 
represents the Poincar\'e dual of ${C \subset M}$.  Equivalently, near
$C$ the gauge field behaves as 
\begin{equation}\label{MONO}
A \,=\, -\frac{\alpha}{2\pi}\,d\vartheta \,+\, \cdots\,,
\end{equation}
where $\vartheta$ is an angular coordinate on the plane transverse to
$C$, located at the origin.  Globally, $A$ has non-trivial monodromy
about any small curve linking $C$ in $M$, and $\SL_\alpha(C)$ is the
reduction to three dimensions of the basic Gukov-Witten
\cite{Gukov:2006jk} surface operator in four dimensions.

Unlike the monopole singularity, the singularity in \eqref{SURFAII}
does not change the degree $m$ of the line-bundle $L$ over $\Sigma$.
As will be useful later, let me give an elementary argument for this
statement.  

We consider an arbitrary configuration for the gauge field
$A$ on ${M = \BR \times \Sigma}$ with the prescribed singularity in
\eqref{SURFAII} at time ${t=0}$, and smooth otherwise.  We will
measure the change in the degree $m$ as the time $t$ runs from
$-\infty$ to $+\infty$ on $M$.  By way of notation, ${\Sigma_\pm
  \subset M}$ will denote the copies of $\Sigma$ at the times ${t =
  \pm \infty}$.  Then the change $\Delta m$ in the degree from ${t =
  -\infty}$ to ${t = +\infty}$ is computed by
\begin{equation}\label{CHGDEG}
\begin{aligned}
\Delta m \,&=\, \int_{\Sigma_+} \! F_A \,-\, \int_{\Sigma_-} \! F_A
\,=\, \int_M \! dF_A\,,\\ 
&=\, -\alpha \int_M \! d\delta_C \,=\, 0\,.
\end{aligned}
\end{equation}
In the first line of \eqref{CHGDEG} we apply Stokes' theorem, and in
the second line we apply the Bianchi identity ${dF_A = 0}$ on the
locus where $A$ is smooth.  Finally, because $C$ is a closed curve on
$\Sigma$, the Poincar\'e dual current $\delta_C$ is also closed,
${d\delta_C = 0}$.  (The same computation would show that ${\Delta m
  \neq 0}$ for the monopole operator, which acts as a localized source
for $dF_A$.)  So as observed following \eqref{LINEAC}, $\SL_\alpha(C)$
must preserve the magnetic label $m$
on the states $|m;\omega\rangle$ in $\SH_\Sigma^{\rm top}$.
  
On the other hand, $\SL_\alpha(C)$ does change the holonomies of $A$
on $\Sigma$, from which the phase factor in \eqref{LINEAC} will be
induced.  To setup the computation, we consider the gauge theory on
${M = \BR \times \Sigma}$, and we suppose that the line operator
$\SL_\alpha(C)$ is inserted on ${\Sigma_0 \equiv \{0\} \times \Sigma}$
in $M$.  We fix an initial flat connection $A_{-}$ on $\Sigma_{-}$.
The operator $\SL_\alpha(C)$ acts as a sudden perturbation to create
the singularity in \eqref{SURFAII} at ${t=0}$, after which we project
$A$ back onto the subspace of harmonic (ie.~flat)
connections.  We then let $A_+$ be the final connection on $\Sigma_+$
which is obtained by subsequent time-evolution.\footnote{Because the
  abelian gauge theory 
  is free, the usual disorder path integral over arbitrary bulk
  configurations for $A$ can be replaced by classical time-evolution.}
We wish to compare the holonomies of $A_+$ to those of $A_-$.  See Figure 1
for a sketch of $M$ as a cylinder over the uniformization of $\Sigma$,
where for concreteness we have drawn $\Sigma$ as a Riemann surface of
genus two.
\begin{figure}[htb!]
\begin{center}
\includegraphics[scale=0.70]{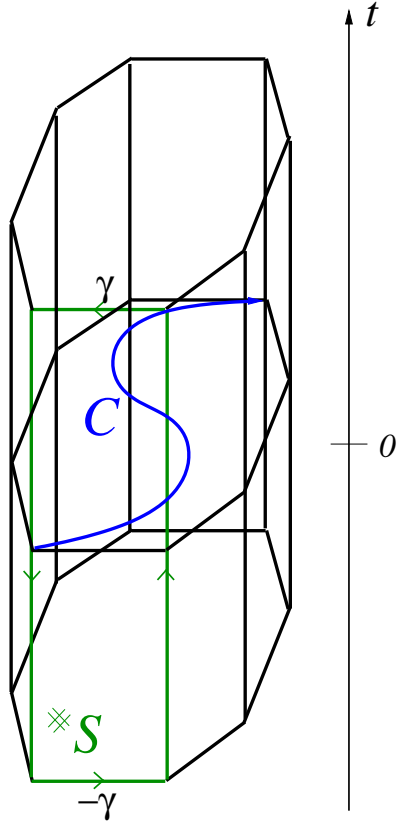}
\caption{$M$ as a cylinder over the uniformization of $\Sigma$.}
\end{center}
\end{figure}

For any closed, oriented curve $\gamma$ on $\Sigma$, we evaluate 
\begin{equation}\label{DELHOL}
\begin{aligned}
\Delta_\gamma A \,&=\, \oint_\gamma A_+ \,-\, \oint_\gamma A_-
\qquad\mod\,\,2\pi\,,\\
&=\, \oint_\gamma A_+ \,+\, \oint_{-\gamma} A_- \qquad\!\!\mod\,\,2\pi\,,
\end{aligned}
\end{equation}
where in the second line we reverse the orientation of $\gamma$ when
integrating $A_-$.  As apparent from Figure
1, we can use Stokes' theorem to evaluate the difference in
\eqref{DELHOL} as an integral over the
cylindrical surface ${S = \BR \times \gamma \subset M}$,
\begin{equation}\label{SURFSI}
\Delta_\gamma A\,=\, \int_{\partial S} A \,=\, \int_S F_A
\qquad\mod\,\,2\pi\,,
\end{equation}
where $S$ is oriented as the rectangle in the figure with boundary 
\begin{equation}\label{SURFSII}
\partial S \,=\, \{+\infty\} \times \gamma \,-\, \{-\infty\} \times
\gamma\,.
\end{equation}
In passing from \eqref{SURFSI} to \eqref{SURFSII}, we note that the
vertical edges of $S$ in Figure 1 are identified after a reversal of
orientation, so they make no contribution to the boundary integral
over $\partial S$ in \eqref{SURFSI}.

We are left to evaluate the integral of the curvature $F_A$ over $S$
in \eqref{SURFSI}.  By assumption, $F_A$ vanishes everywhere on $S$
except for the explicit curvature singularity \eqref{SURFAII} created
at ${t=0}$ along $C$.  Consequently,
\begin{equation}\label{DELHOLV}
\Delta_\gamma A \,=\, -\alpha \int_S \delta_C \,=\, \alpha
\oint_\gamma\, [C]^\vee \qquad\mod\,\,2\pi\,,
\end{equation}
where ${[C]^\vee \in H^1(\Sigma;\BZ)}$ is a harmonic representative
for the dual of the homology class of $C$.  The
flip of sign in the second equality arises from due care with orientations.

Because the curve $\gamma$ is arbitrary, $\SL_\alpha(C)$ must act
classically by the shift 
\begin{equation}\label{APM}
A_+ \,=\, A_- \,+\, \alpha \, [C]^\vee \quad\mod\,\, 2\pi\BL\,.
\end{equation}
We recall from \eqref{EAP} that the wavefunction $\Psi_\omega$
evaluated on a harmonic connection $A$ of degree $m$ is given by 
\begin{equation}
\Psi_\omega(A) \,=\, \exp{\!\left[i\int_\Sigma \big(A \,-\,
      m\,\widehat{A}\,\big)\^\omega\right]}\,,
\end{equation}
Under the shift \eqref{APM} induced by $\SL_\alpha(C)$, we see that
$\Psi_\omega$ transforms by
\begin{equation}\label{LAC}
\begin{aligned}
\SL_\alpha(C)\!\left[\Psi_\omega(A)\right] \,&=\, \Psi_\omega\!\left(A
  \,+\, \alpha \, [C]^\vee\right)\,,\\
&=\exp{\!\left[i \, \alpha \int_\Sigma [C]^\vee \^ \omega\right]}
\cdot \Psi_\omega\big(A\big)\,,\\ 
&=\exp{\!\left[i \, \alpha \oint_C \omega\right]} \cdot
\Psi_\omega\big(A\big)\,.
\end{aligned}
\end{equation}
exactly as in \eqref{LINEAC}.

I briefly mention two other ways to understand the formula in
\eqref{LAC} from gauge theory.

So far we have introduced two disorder operators for the gauge field
$A$, namely, the monopole operator $\SV_k(\sigma)$ and the loop
operator $\SL_\alpha(C)$.  These two operators are not unrelated.  Let
us again consider a monopole/anti-monopole pair
${\SV_k(\sigma) \circ \SV_{-k}(\sigma_0)}$ on $\Sigma$.  If ${\sigma =
  \sigma_0}$, this composite operator is the identity, but for
${\sigma \neq \sigma_0}$, the operator acts on states in
$\SH_\Sigma^{\rm top}$ with the non-trivial phase 
\begin{equation}\label{VERLD}
\SO_k(\sigma,\sigma_0)|m;\omega\rangle \,=\, 
\Big[\SV_k(\sigma) \circ \SV_{-k}(\sigma_0)\Big]|m;\omega\rangle \,=\,
\exp{\!\left( 2 \pi i \, k \int^\sigma_{\sigma_0} \omega\right)}
|m;\omega\rangle\,.
\end{equation}

Using $\SO_k(\sigma,\sigma_0)$, we can try to make a new
operator \'a la Verlinde \cite{Verlinde:1988} from the 
monodromy action by ${\SO_k(\sigma,\sigma_0)}$ on
$\SH_\Sigma^{\rm top}$ as the point $\sigma$ is moved adiabatically
around a closed curve ${C \subset \Sigma}$ based at $\sigma_0$.  When
$k$ is integral, the induced phase in \eqref{VERLD} is trivial, and
the Verlinde operator acts as the identity.  However, when ${k \neq 0
  \,\mod\,\BZ}$ is allowed to be fractional, the Verlinde operator
constructed from ${\SO_k(\sigma,\sigma_0)}$ is non-trivial and acts by
precisely the phase in \eqref{LAC}, provided we set 
\begin{equation}\label{FRACQ}
\alpha \,=\, 2 \pi k \,\in\, \BR/2\pi\BZ\,.
\end{equation}
So the line operator $\SL_\alpha(C)$ can be interpreted as the
Verlinde operator associated to transport of a monopole/anti-monopole
pair with non-integral magnetic charge.  See \cite{Gukov:2006jk}
for a somewhat different explanation of the same effect in
four-dimensional gauge theory.  This relation can also be understood
directly from the order-type expressions for
${\SV_k(\sigma)\circ\SV_{-k}(\sigma_0)}$ and $\SL_\alpha(C)$ in terms
of the periodic scalar field $\phi$.

Alternatively, the action by the line operator $\SL_\alpha(C)$ can be
understood using the Lagrangian formalism, the focus of
\cite{BeasleyI:2014}.  For simplicity in the following discussion, we
assume that the topological parameters $\theta$ and $\upbeta$ from
Section \ref{Maxwell} are both set to zero.  In the Lagrangian
formalism, the inner-product of states ${\langle
  m;\omega'|\SL_\alpha(C)|m;\omega\rangle}$ for some
${\omega,\omega'\in\BL}$ is computed by the path
integral 
\begin{equation}\label{PIRLA}
\begin{aligned}
&\big\langle m;\omega'|\SL_\alpha(C)|m;\omega\big\rangle \,=\, \\
&\frac{1}{\Vol(\CG)} \int_{\Pic_m(\Sigma) \times \CA_m \times
  \Pic_m(\Sigma)} \mskip-45mu \CD A_+ 
\, \CD A \, \CD A_- \,\, \bar\Psi_{\omega'}(A_+) \,
\exp{\!\left[{\frac{i}{4\pi e^2}}\int_M \CF_A \^ \* \CF_A
  \right]} \,  \Psi_\omega(A_-)\,,
\end{aligned}
\end{equation}
with modified curvature
\begin{equation}\label{WTFA}
\CF_A \,=\, F_A \,+\, \alpha \, \delta_C\,.
\end{equation}
Here $A_\pm$ denote the boundary values for the gauge field at ${t =
  \pm \infty}$, and the path integral ranges over the affine space $\CA_m$ of
connections on the $U(1)$-bundle with degree $m$ on ${M = \BR \times
  \Sigma}$.  We also integrate over the boundary values of $A$ with
weights given by the wavefunctions $\Psi_\omega$ and $\Psi_{\omega'}$
as in \eqref{EAP}.  Finally, the term proportional to $\delta_C$ in \eqref{WTFA} 
enforces the condition that $F_A$ have the singular behavior in
\eqref{SURFAII}.  For a more thorough discussion of the latter remark,
see Section $5.2$ in \cite{BeasleyI:2014}.

The modified action in \eqref{PIRLA} can be expanded in
terms of $F_A$ as  
\begin{equation}\label{PIRLASY}
\int_M \CF_A \^ \* \CF_A \,=\, \int_M F_A\^\*F_A \,+\, 2 \alpha
\int_M \! \delta_C \^ \* F_A \,+\, c_0\,.
\end{equation}
Here $c_0$ is a formally divergent constant arising from the
norm-square of $\delta_C$, which we shall ignore.  By comparison of 
\eqref{PIRLASY} to the standard Maxwell action, $\SL_\alpha(C)$ can 
be identified semi-classically with the operator
\begin{equation}\label{PIRLAII}
\SL_\alpha(C) \,=\, \exp{\!\left[\frac{i\,\alpha}{2\pi e^2} \int_M \!
    \delta_C \^ \* F_A\right]} \,=\, \exp{\!\left[\frac{i\,\alpha}{2\pi
      e^2} \oint_C \*_\Sigma E_A\right]}\,,
\end{equation}
where I note in the second equality that only the electric component
of $F_A$ contributes to the integral over the spacelike curve ${C
  \subset \Sigma}$, and $\*_\Sigma$ indicates the two-dimensional
Hodge operator on $\Sigma$.  Upon 
quantization as in \eqref{DOVDAE},\footnote{The Coulomb-gauge smearing term in
  \eqref{DOVDAE} can be ignored when we restrict to the topological subspace
  $\SH_\Sigma^{\rm top}$.}
\begin{equation}
\frac{1}{2\pi e^2}\,\*_\Sigma E_A(w) \,=\,
-i\,\vol_\Sigma\cdot\frac{\delta}{\delta A(w)}\,,
\end{equation}
so the loop operator becomes 
\begin{equation}\label{PIRLAIII}
\SL_\alpha(C) \,=\, \exp{\!\left[\alpha \oint_C \vol_\Sigma \cdot 
    \frac{\delta}{\delta A}\right]}\,.
\end{equation}
Manifestly, $\SL_\alpha(C)$ acts upon any wavefunction
$\Psi_\omega(A)$ by the shift ${A \mapsto A + \alpha\,[C]^\vee}$
appearing in the first line of \eqref{LAC}.

Although we began with a disorder characterization of the loop
operator in gauge theory, the classical description for $\SL_\alpha(C)$
in \eqref{PIRLAII} amounts to an order expression for the same
operator.  A quantum operator may admit distinct classical
descriptions, so there is no contradiction here.  See for instance
Section $4.1$ of \cite{Beasley:2009mb} for an analogous disorder
presentation of the usual Wilson loop operator in Chern-Simons gauge theory.

\medskip\noindent{\sl Wilson Loops}\medskip

We are left to consider the action on $\SH_\Sigma^{\rm top}$ of the
Wilson loop operator $\SW_n(C)$.  In terms of the
gauge field, the Wilson loop operator acts simply by
multiplication in the topological sector labelled by the degree $m$,
\begin{equation}\label{WLOOP}
\begin{aligned}
\SW_n(C)\cdot\Psi_\omega(A) \,&=\, \exp{\!\left(i \, n \oint_C
    A\right)} \cdot \exp{\!\left[i\int_\Sigma \big(A - m\,
    \widehat{A}\big)\^\omega\right]}\,,\\
&=\, \exp{\!\left(i \, n \int_\Sigma [C]^\vee \^ A\right)} \cdot
\exp{\!\left[i\int_\Sigma \big(A - m \, \widehat{A}\big)\^\omega\right]}\,,\\
&=\, \exp{\!\left(i\,m\,n \oint_C \widehat{A}\right)} \cdot
\Psi_{\omega \,-\, n \, [C]^\vee}(A)\,,
\end{aligned}
\end{equation}
where again ${[C]^\vee \in H^1(\Sigma;\BZ)}$ is the Poincar\'e dual of the
curve ${C \subset \Sigma}$.  In passing to the second line of
\eqref{WLOOP}, we recall that $A$ is a harmonic connection with
expansion \eqref{SOMEA} for wavefunctions in $\SH_\Sigma^{\rm top}$.
As a result, 
\begin{equation}\label{WLOOPII}
\SW_n(C) \big|m;\omega\big\rangle \,=\,
  \exp{\!\left(i\,m\,n \oint_C \widehat{A}\right)} \big|m;\omega -
  n \, [C]^\vee\big\rangle\,.
\end{equation}
Clearly, integrality of $n$ is necessary whenever
the homology class ${[C] \neq 0}$ is non-trivial, else the Wilson loop
operator does not act in a well-defined way on $\SH_\Sigma^{\rm top}$.

The phase factor in \eqref{WLOOPII} depends upon the fiducial harmonic
connection $\widehat{A}$ on the degree-one line-bundle
${L=\CO_\Sigma(\sigma_0)}$ over $\Sigma$.  Because $\widehat{A}$ is not
flat, this phase is not invariant under deformations of the curve $C$.  For
instance, even when ${C = \partial D}$ is trivial in homology, the Wilson
loop operator still acts non-trivially on the state
$|m;\omega\rangle$,
\begin{equation}\label{WLOOPA}
\begin{aligned}
\SW_\nu(C) \big|m;\omega\big\rangle \,&=\, \exp{\!\left(i \, m \,\nu
    \int_D \widehat{F}_A\right)} \big|m;\omega\big\rangle\,,\\
&=\, \exp{\!\left[2\pi i \, m \, \nu
    \!\left(\frac{\vol_\Sigma(D)}{\ell^2}\right)\right]}
\big|m;\omega\big\rangle\,,\qquad\qquad C \,=\, \partial D\,.
\end{aligned}
\end{equation}
In passing to the second line, we use the formula for $\widehat{F}_m$
in \eqref{FHAT}, and we let $\vol_\Sigma(D)$
be the volume of $D$ in the given metric on $\Sigma$.  If $\nu$
is integral, the phase in \eqref{WLOOPA} does not depend on whether
$D$ or ${D' = \Sigma - D}$ is chosen to bound $C$.  Otherwise, for
arbitrary real values ${\nu \in \BR}$, the orientation of $C$
uniquely fixes the bounding two-cycle $D$ with compatible
orientation, so that the action of $\SW_\nu(C)$ is 
well-defined.\footnote{As observed in Section $5.2$ of
  \cite{BeasleyI:2014}, the analogous definition of $\SW_\nu(C)$ for
  null-homologous curves $C$ in a three-manifold $M$ generally
  does depend upon an extra discrete choice of a relative class in
  $H_2(M,C)$ for the bounding Seifert surface.}

As usual, we now wish to understand the results in \eqref{WLOOPII}
and \eqref{WLOOPA} dually in terms of the periodic scalar field
$\phi$.  Like the previous disorder description for the loop
operator $\SL_\alpha(C)$, the Wilson loop operator $\SW_n(C)$ will
act on $\phi$ by creating a singularity along $C$ such that $\phi$
winds by ${2\pi n}$ when traversing any small circle which links ${C
  \subset M = \BR \times \Sigma}$.

The effective shift of ${\omega \in H^1(\Sigma;\BZ)}$ in
\eqref{WLOOPII} can be understood dually in close correspondence to
the shift \eqref{APM} induced by the vortex loop operator
$\SL_\alpha(C)$ on the gauge field $A$.  Classically, $\omega$ is
interpreted as the winding-number of $\phi$, with ${\omega =
  [d\phi/2\pi]}$.  So long as $\phi$ is a smooth map to the circle,
then ${d[d\phi] = 0}$.  However, in the background of the
Wilson loop operator $\SW_n(C)$, we replace the smooth map $\phi$ by
a section $\wt\phi$ of a non-trivial $S^1$-bundle over the complement
${M^o = M - C}$, such that 
\begin{equation}
d[d\wt\phi] \,=\, d(d\phi + B) \,=\, F_B \,=\, 2\pi n\,\delta_C\,.
\end{equation}
For a path integral justification of the statement above, I refer the
interested reader to the end of Section $5.2$ in
\cite{BeasleyI:2014}.  

By exactly the same computation as in
\eqref{DELHOL}, \eqref{SURFSI}, and \eqref{DELHOLV}, we evaluate the
change due to the insertion of $\SW_n(C)$ in the winding-number of
$\phi$ around an arbitrary closed curve ${\gamma\subset \Sigma}$ as
\begin{equation}\label{DELGAM}
\begin{aligned}
\Delta_\gamma\omega \,&=\, \oint_{\gamma} \frac{d\phi_+}{2\pi}
\,-\, \oint_{\gamma} \frac{d\phi_-}{2\pi}\,,\\ 
&=\,\int_S d\!\left[\frac{d\wt\phi}{2\pi}\right],\qquad\qquad\qquad S
\,=\, \BR \times \gamma\,,\\
&=\, n \int_S \delta_C \,=\, -n \, \oint_\gamma
[C]^\vee\,.
\end{aligned}
\end{equation}
reproducing the shift in \eqref{WLOOPII}.  

The non-topological, $m$-dependent phase in \eqref{WLOOPII} is
slightly more subtle.  For simplicity, we will reproduce this phase
only in the special case that ${C = \partial D}$ is
homologically-trivial, as assumed in \eqref{WLOOPA}.  Then 
\begin{equation}\label{WLOOPF} 
\SW_\nu(C) \,=\, \exp{\!\left[i \, \nu \int_D
    F_A\right]}\,,\qquad\qquad\qquad C \,=\, \partial D\,.
\end{equation}
As the ur-statement of abelian duality, discussed in the Introduction
to \cite{BeasleyI:2014}, we have the correspondence
\begin{equation}
F_A = e^2\,\*d\phi\,.
\end{equation}
Hence the classical description of the Wilson loop operator in terms of
$\phi$ must be 
\begin{equation}
\SW_\nu(C) \,=\, \exp{\!\left[i \, \nu \, e^2 \int_D
    \vol_\Sigma\cdot \partial_t\phi\right]}\,.
\end{equation}
Upon quantization, we apply the functional identification in
\eqref{DOVDPHI} with ${\theta = 0}$ to rewrite $\SW_\nu(C)$ as the
operator 
\begin{equation}\label{CLWNCU}
\SW_\nu(C) \,=\, \exp{\!\left[2\pi\nu \int_D
    \vol_\Sigma\cdot\frac{\delta}{\delta\phi}\right]}\,.
\end{equation}
Hence $\SW_\nu(C)$ acts upon any wavefunction $\Psi_m(\phi)$ by the
shift $\phi \mapsto \phi + 2\pi \nu \, \vol_\Sigma(D)/\ell^2$.

According to our previous results in \eqref{EIGENOM} and
\eqref{FOURIER}, the Fourier wavefunction $\Psi_m(\phi)$ which describes the
Fock state $|m;\omega\rangle$ is given explicitly by 
\begin{equation}
\Psi_m(\phi) \,=\, \exp{\!\left[i\,\frac{m}{\ell^2}\int_\Sigma
    \vol_\Sigma\cdot\left(\phi - \Phi_\omega\right)\right]}\,.
\end{equation}
Immediately, the action by the operator in \eqref{CLWNCU} on this
wavefunction produces the geometric phase in the second line of
\eqref{WLOOPA}.  

To summarize, the Wilson loop $\SW_n(C)$ and the vortex loop $\SL_\alpha(C)$
play dual roles.  When expressed in terms of
the gauge field, $\SW_n(C)$ acts classically by multiplication on
any state $|m;\omega\rangle$.  But when expressed in terms of the
scalar field, $\SW_n(C)$ acts quantum-mechanically as the
differential (or shift) operator in \eqref{CLWNCU}.  Conversely, the vortex loop
$\SL_\alpha(C)$ acts classically by multiplication when written in
terms of $\phi$, but quantum-mechanically as the differential (or
shift) operator in \eqref{PIRLAIII} when written in terms of $A$.

\subsection{Wilson-'t Hooft Commutation Relations}\label{Heisenberg}

Finally, let us examine the commutation relations between the operators
$\SV_k(\sigma)$, $\SL_\alpha(C)$, and $\SW_n(C)$, all acting on the 
topological Hilbert space $\SH_\Sigma^{\rm top}$.  The idea of
examining these commutators goes back to 't Hooft, and we will find a
holomorphic refinement of the classic results in \cite{tHooft:1977hy}.

Collecting our previous formulas in \eqref{VSTOP}, \eqref{LINEAC}, and
\eqref{WLOOPII} we explicitly present the action of the operators on
the Fock vacua $|m;\omega\rangle$ as   
\begin{equation}\label{OPACT}
\begin{aligned}
\SV_k(\sigma) |m;\omega\rangle \,&=\, \exp{\!\left(2 \pi i  \, k
    \int^\sigma_{\sigma_0} \omega\right)} |m+k;\omega\rangle\,,\\ 
\SL_\alpha(C) |m;\omega\rangle \,&=\, \exp{\!\left(i \, \alpha
    \oint_C \omega\right)} |m;\omega\rangle\,,\\ 
\SW_n(C) \big|m;\omega\big\rangle \,&=\, \exp{\!\left(i\,m\,n
    \oint_C \widehat{A}\right)} \big|m;\omega - n \,
[C]^\vee\big\rangle\,.
\end{aligned}
\end{equation}
Clearly for all pairs ${\sigma, \sigma' \in \Sigma}$ and ${C, C' \subset \Sigma}$,
\begin{equation}\label{COMMI}
\Big[\SV_k(\sigma),\, \SV_{k'}(\sigma')\Big] \,=\, \Big[\SL_\alpha(C),\,
\SL_{\alpha'}(C')\Big] \,=\, \Big[\SW_n(C),\,
\SW_{n'}(C')\Big] \,=\, 0\,.
\end{equation}
Also, 
\begin{equation}\label{COMMII}
\Big[\SV_k(\sigma),\, \SL_\alpha(C)\Big] \,=\, 0\,,
\end{equation}
as follows directly from the elementary, order-type description of
both the vertex and the homological loop operators in terms of the periodic
scalar field $\phi$.

On the other hand, the loop operators $\SL_\alpha(C)$ and $\SW_n(C)$
do not commute.  Instead, the composition satisfies
\begin{equation}\label{COMMIII}
\SL_\alpha(C) \circ \SW_n(C') \,=\, \exp{\!\left[-i
      \, \alpha \, n \left(C \cdot C'\right)\right]} \; \SW_n(C')
  \circ \SL_\alpha(C)\,,
\end{equation}
where 
\begin{equation}
C \cdot C' \,=\, \oint_{C} \left[C'\right]{}^{\!\vee} \,\in\,\BZ\,.
\end{equation}
Equivalently, ${C \cdot C'}$ is the topological intersection number of the
curves ${C, C' \subset \Sigma}$.  Because $n$ and ${C \cdot C'}$ are
integers, the phase in \eqref{COMMIII} only depends upon the value of
$\alpha$ modulo $2\pi$, consistent with its angular nature.  When $C'$
is trivial in homology, the charge $n$ of the Wilson loop can be replaced by an arbitrary real
parameter $\nu$.  In this special case, the phase in
\eqref{COMMIII} remains well-defined, since ${C\cdot C' = 0}$.

Of course, the commutator in \eqref{COMMIII} appears in direct analogy
to the celebrated commutation relation for Wilson
and 't Hooft operators in four-dimensional abelian gauge theory, for
which the corresponding phase is proportional to the linking number of
the curves $C$ and $C'$ in $\BR^3$.  See \S $10.2$ of \cite{Deligne}
for a review of this story in four dimensions. 

Though more or less obvious, the non-trivial commutation relation in
\eqref{COMMIII} has an interesting consequence, because it implies that the
monopole operator $\SV_k(\sigma)$ and the Wilson loop operator
$\SW_n(C)$ similarly fail to commute.  As one can check directly, 
\begin{equation}\label{COMMIV}
\SV_k(\sigma) \circ \SW_n(C) \,=\, \exp{\!\left(-2\pi i \, k \, n
    \int_{\sigma_0}^\sigma \left[C\right]{}^{\!\vee}\right)} \exp{\!\left(-i \,
    k \, n \oint_C \widehat{A}\right)} \; \SW_n(C) \circ \SV_k(\sigma)\,.
\end{equation}
To make sense of \eqref{COMMIV}, we must work with a definite,
harmonic representative for the cohomology class $[C]^\vee$ which is
Poincar\'e dual to $[C]$.  Otherwise, absent a definite
representative, the value of the line integral from $\sigma_0$ 
to $\sigma$ in \eqref{COMMIV} would be ambiguous.

One potentially unsettling feature of the commutation relation in
\eqref{COMMIV} is that the phase on the right-hand side appears to depend upon the
auxiliary choices of the basepoint ${\sigma_0 \in \Sigma}$ and the harmonic
connection $\widehat{A}$.  These choices enter the definition of the
states $|m;\omega\rangle$, but they do not enter the
intrinsic definitions of the operators $\SV_k(\sigma)$ and
$\SW_n(C)$ themselves and hence should not enter the
commutator.\footnote{I thank Marcus Benna for emphasizing this question 
  to me.}  

Actually, the situation is slightly better than it first appears,
since the fiducial connection $\widehat{A}$ is itself 
determined by the choice of $\sigma_0$.  We recall that
${\widehat{A}}$ is defined as the unique harmonic connection 
compatible with the holomorphic structure on $\CO_\Sigma(\sigma_0)$.
As we now demonstrate, the explicit  
dependence on $\sigma_0$ in the first phase factor of
\eqref{COMMIV} exactly cancels against the implicit dependence of
${\widehat{A} \equiv \widehat{A}_{\sigma_0}}$ on $\sigma_0$ in the second phase factor.  

We begin by introducing another harmonic connection
$\widehat{A}_\sigma$, associated to the degree-one holomorphic
line-bundle $\CO_\Sigma(\sigma)$.   As harmonic connections, both $\widehat{A}_\sigma$
and $\widehat{A}_{\sigma_0}$ have the same curvature, proportional
to the Riemannian volume form on $\Sigma$, so the difference   
${\widehat{A}_\sigma - \widehat{A}_{\sigma_0}}$ is a closed one-form.
For the first phase factor in \eqref{COMMIV}, the 
classical Abel-Jacobi theory now provides the very beautiful
reciprocity relation 
\begin{equation}\label{ABJac} 
\exp{\!\left(-2\pi i \, k \, n \int_{\sigma_0}^\sigma
      \left[C\right]{}^{\!\vee}\right)} \,=\, \exp{\!\left[-i \, k \,
      n \oint_C \left(\widehat{A}_\sigma - \widehat{A}_{\sigma_0}\right)\right]}\,.
\end{equation}
See Ch.\,$2.2$ of \cite{Griffiths:78} for a textbook reference on such
reciprocity laws.

Substituting \eqref{ABJac} into \eqref{COMMIV}, we obtain a completely 
intrinsic reformulation of the commutation relation between the
monopole operator and the Wilson loop, 
\begin{equation}\label{COMMP}
\SV_k(\sigma) \circ \SW_n(C) \,=\, \exp{\!\left(-i \, k \, n
    \oint_C \widehat{A}_\sigma\right)} \; \SW_n(C) \circ 
  \SV_k(\sigma)\,,
\end{equation}
with no dependence on the arbitrary choice of the basepoint $\sigma_0$.  We
emphasize that the commutation relation in \eqref{COMMP} {\sl does} depend
on the particular curve ${C \subset \Sigma}$, not merely the homology
class $[C]$, because $\widehat{A}_\sigma$ is not flat.  Moreover, the
commutator depends holomorphically on the point at
which the monopole operator is inserted, through the dependence of
$\widehat{A}_\sigma$ on $\sigma$.

The commutation relation in \eqref{COMMP} can be
understood directly in terms of either the gauge field $A$ or the
periodic scalar field $\phi$.  Via the Hecke modification in \eqref{HECKE}, the monopole operator
$\SV_k(\sigma)$ induces the shift 
\begin{equation}\label{SHFTA}
 A \,\longmapsto\, A \,+\, k\,\widehat{A}_\sigma\,.
\end{equation}
On the other hand, the Wilson loop operator $\SW_n(C)$ acts
multiplicatively as in \eqref{WLOOP}, from which \eqref{COMMP} follows.  
Alternatively in terms of $\phi$, the vertex operator $\SV_k(\sigma)$
acts multiplicatively, and the Wilson loop operator
$\SW_\nu(C)$ for homologically-trivial ${C = \partial D}$
acts as the shift operator in \eqref{CLWNCU}, from which \eqref{COMMP}
again follows.

To gain a final bit of additional insight into the meaning of the
Wilson-'t Hooft commutation relation, let us
consider the commutation relation of 
$\SW_n(C')$ with the composite operator ${\SO_k(\sigma,\sigma_0) \equiv \SV_k(\sigma) \circ
  \SV_{-k}(\sigma_0)}$ which has vanishing monopole charge.  As an immediate
consequence of \eqref{COMMIV},
\begin{equation}\label{COMMV} 
\SO_k(\sigma,\sigma_0) \circ \SW_n(C') \,=\,
\exp{\!\left(-2\pi i \, k \, n \int_{\sigma_0}^\sigma
    \left[C'\right]{}^{\!\vee}\right)}\; \SW_n(C') \circ \SO_k(\sigma,\sigma_0)\,.
\end{equation}
But we have already identified $\SL_\alpha(C)$ as the Verlinde
operator derived from $\SO_k(\sigma,\sigma_0)$, describing the creation
and subsequent transport of a monopole/anti-monopole pair with
fractional magnetic charge ${\alpha = 2\pi k}$.  As $\sigma$ is
transported adiabatically around a curve $C$ based at $\sigma_0$, the
commutation relation in \eqref{COMMV} reproduces the topological
commutation relation of loop operators in \eqref{COMMIII}.

\bigskip
\noindent{\bf Acknowledgments}\smallskip

I take pleasure in thanking Marcus Benna, Martin Ro\v cek, and
A.J.\,Tolland for conversations on these and related matters.  I
acknowledge support from the Simons Center for Geometry and Physics,
where a portion of this work was completed.

\end{onehalfspace}
\bibliographystyle{unsrt} 

\begin{thebibliography}{99}
\begin{raggedright}

\bibitem{Beasley:2009mb}
C.~Beasley, ``Localization for Wilson Loops in Chern-Simons Theory,''
Adv.\ Theor.\ Math.\ Phys.\ {\bf 17} (2013) 1--240,
\href{http://arXiv.org/abs/arXiv:0911.2687}{\tt 
  arXiv:0911.2687\,[hep-th]}.

\bibitem{BeasleyI:2014}
C.~Beasley, ``Global Aspects of Abelian Duality in Dimension Three,''
\href{http://arxiv.org/abs/1405.2123}{\ttfamily arXiv:1405.2123\,[hep-th]}.

\bibitem{Borokhov:2002ib} 
V.~Borokhov, A.~Kapustin, and X.-k.~Wu,
``Topological Disorder Operators in Three-Dimensional Conformal Field
Theory,'' JHEP {\bf 0211} (2002) 049, \href{http://arxiv.org/abs/hep-th/0206054}{{\ttfamily
    hep-th/0206054}}.

\bibitem{Borokhov:2002cg}
V.~Borokhov, A.~Kapustin, and X.-k.~Wu,
``Monopole Operators and Mirror Symmetry in Three-Dimensions,'' JHEP
{\bf 0212} (2002) 044, \href{http://arxiv.org/abs/hep-th/0207074}{{\ttfamily
    hep-th/0207074}}.

\bibitem{BrodaWG}
B.~Broda and G.~Duniec, ``Abelian Duality in Three-Dimensions,''
Phys.\ Rev.\ D {\bf 70} (2004)
107702, \href{http://arxiv.org/abs/hep-th/0210151}{{\ttfamily hep-th/0210151}}.

\bibitem{Buscher:1985}
T.~Buscher, ``Quantum Corrections and Extended Supersymmetry in New
$\sigma$-Models,'' Phys.\ Lett.\ B {\bf 159} (1985) 127--130.

\bibitem{BuscherSK}
T.~Buscher, ``A Symmetry of the String Background Field Equations,''
Phys.\ Lett.\ B {\bf 194} (1987) 59--62.

\bibitem{BuscherQJ}
T.~Buscher, ``Path Integral Derivation of Quantum Duality in Nonlinear
Sigma Models,'' Phys.\ Lett.\ B {\bf 201} (1988) 466--472.

\bibitem{Drukker:2012sr}
N.~Drukker, T.~Okuda, and F.~Passerini, ``Exact Results for Vortex
Loop Operators in 3d Supersymmetric Theories,''
\href{http://arxiv.org/abs/1211.3409}{\ttfamily arXiv:1211.3409\,[hep-th]}. 

\bibitem{Freed:2006ya}
D.~Freed, G.~Moore, and G.~Segal, ``The Uncertainty of Fluxes,''
  Commun.\ Math.\ Phys.\  {\bf 271} (2007) 247--274,
  \href{http://arxiv.org/abs/hep-th/0605198}{\ttfamily hep-th/0605198}.

\bibitem{Freed:2006yc}
D.~Freed, G.~Moore, and G.~Segal, ``Heisenberg Groups and
Noncommutative Fluxes,'' Annals Phys.\  {\bf 322} (2007) 236--285,
\href{http://arxiv.org/abs/hep-th/0605200}{{\ttfamily
    hep-th/0605200}}.

\bibitem{Friedman:98}
R.~Friedman, {\it Algebraic Surfaces and Holomorphic Vector Bundles},
Springer-Verlag, New York, 1998.

\bibitem{Giveon:1994fu}
A.~Giveon, M.~Porrati, and E.~Rabinovici, ``Target Space Duality in
String Theory,'' Phys.\ Rept.\  {\bf 244} (1994) 77--202,
\href{http://arXiv.org/abs/hep-th/9401139}{\ttfamily hep-th/9401139}.

\bibitem{Griffiths:78}
P.~Griffiths and J.~Harris, {\it Principles of Algebraic Geometry},
John Wiley and Sons, Inc., New York, 1978.

\bibitem{Gukov:2006jk} 
S.~Gukov and E.~Witten, ``Gauge Theory, Ramification, and the
Geometric Langlands Program,'' in {\it Current Developments in
Mathematics, 2006}, Ed. by D.~Jerison et. al., pp. 35--180, Int.\
Press, Somerville, Massachusetts, 2008,
\href{http://arXiv.org/abs/hep-th/0612073}{\ttfamily hep-th/0612073}.

\bibitem{tHooft:1977hy}
G.~'t Hooft, ``On the Phase Transition Towards Permanent Quark
Confinement,'' Nucl.\ Phys.\ B {\bf 138} (1978) 1--25.

\bibitem{Intriligator:2013lca}
K.~Intriligator and N.~Seiberg, ``Aspects of 3d ${\CN=2}$
Chern-Simons-Matter Theories,'' JHEP {\bf 1307} (2013) 079,
\href{http://arXiv.org/abs/1305.1633}{\ttfamily arXiv:1305.1633\,[hep-th]}.

\bibitem{Kapustin:2006pk}
A.~Kapustin and E.~Witten, ``Electric-Magnetic Duality and the
Geometric Langlands Program,'' Commun.\ Number Theory Phys.\ {\bf 1}
(2007) 1--236, \href{http://arXiv.org/abs/hep-th/0604151}{\tt
  hep-th/0604151}.

\bibitem{Kapustin:2009av}
A.~Kapustin and M.~Tikhonov, ``Abelian Duality, Walls, and Boundary
Conditions in Diverse Dimensions,'' JHEP {\bf 0911} (2009) 006,\\
\href{http://arxiv.org/abs/arXiv:0904.0840}{\ttfamily
  arXiv:0904.0840\,[hep-th]}.

\bibitem{Kapustin:2012iw}
A.~Kapustin, B.~Willett, and I.~Yaakov, ``Exact Results for
Supersymmetric Abelian Vortex Loops in 2+1 Dimensions,''  JHEP {\bf
  1306} (2013) 099, \href{http://arXiv.org/abs/1211.2861}{\ttfamily
  arXiv:1211.2861\,[hep-th]}.

\bibitem{ProdanovJY}
E.~~Prodanov and S.~Sen, ``Abelian Duality,''
Phys.\ Rev.\ D {\bf 62} (2000)
045009, \href{http://arxiv.org/abs/hep-th/9906143}{{\ttfamily
    hep-th/9906143}}.

\bibitem{RocekPS}
M.~Ro\v cek and E.~Verlinde, ``Duality, Quotients, and Currents,''
Nucl.\ Phys.\ B {\bf 373} (1992) 630--646,
\href{http://arxiv.org/abs/hep-th/9110053}{\ttfamily hep-th/9110053}.

\bibitem{Verlinde:1988}
E.~Verlinde, ``Fusion Rules and Modular Transformations in 2D
Conformal Field Theory,'' Nucl.\ Phys.\ B {\bf 300} (1988) 360--376.

\bibitem{Deligne}
E.~Witten, ``Dynamics of Quantum Field Theory,'' in {\it Quantum Fields
and Strings:\ A Course for Mathematicians, Vol. 2}, Ed. by P. Deligne et al.,
American Mathematical Society, Providence, Rhode Island, 1999.

\bibitem{Witten:2009mh} 
E.~Witten, ``Geometric Langlands and the Equations of Nahm and
Bogomolny,'' \href{http://arXiv.org/abs/arXiv:0905.4795}{{\tt
arXiv:0905.4795\,[hep-th]}}.

\end{raggedright}
\end{thebibliography}

\end{document}